\documentclass[
 preprint, 
 superscriptaddress,
 amsmath,amssymb,
 aps, 
 physrev,
]{revtex4-2}

\usepackage{graphicx}
\usepackage{dcolumn}
\usepackage{bm}
\usepackage{soul}
\usepackage{xcolor}

\begin{document}

\title{\textbf{Anomalous Reversal of Stability in Mo-containing Oxides: \\A Difficult Case Exhibiting Sensitivity to DFT+U and Distortion} 
}%

\author{Tzu-chen Liu}
\affiliation{Department of Materials Science and Engineering, Northwestern University, Evanston, IL 60208, USA
}

\author{Dale Gaines II}
\affiliation{Department of Materials Science and Engineering, Northwestern University, Evanston, IL 60208, USA
}

\author{Hyungjun Kim}
\affiliation{Department of Materials Science and Engineering, Northwestern University, Evanston, IL 60208, USA
}

\author{Adolfo Salgado-Casanova}
\affiliation{Department of Materials Science and Engineering, Northwestern University, Evanston, IL 60208, USA
}

\author{Steven B. Torrisi}
\affiliation{Energy \& Materials Division, Toyota Research Institue, Los Altos, CA 94022, USA}

\author{Chris Wolverton}
 \email{c-wolverton@northwestern.edu }
\affiliation{Department of Materials Science and Engineering, Northwestern University, Evanston, IL 60208, USA
}

\date{\today}

\begin{abstract}
Accurate predictions of the properties of transition metal oxides using density functional theory (DFT) calculations are essential for the computational design of energy materials. In this work, we investigate the anomalous reversal of the stability of structural distortions (where distorted structures go from being energetically favorable to sharply unfavorable relative to undistorted ones) induced by DFT+U on Mo d-orbitals in layered AMoO$_2$ (A = Li, Na, K) and rutile-like MoO$_2$. We highlight the significant impact of varying U$_{\text{eff}}$ values on the structural stability, convex hull, and thermodynamic stability predictions, noting that deviations can reach up to the order of 100 meV/atom across these energetic quantities. We find the transitions in stability are coincident with changes in the electron localization and magnetic behavior. The anomalous reversal persists across PBE, r$^2$SCAN functionals, and also with vdW-dispersion energy corrections (PBE+D3). In Mo-containing oxide systems, high U$_{\text{eff}}$ leads to inaccurate descriptions of physical quantities and structural relaxations under artificial symmetry constraints, as demonstrated by the phonon band structures, the Heyd-Scuseria-Ernzerhof (HSE06) hybrid functional results, and comparisons with experimental structural data. We conclude that high U$_{\text{eff}}$ values (around 4 eV and above, depending on the specific structures and compositions) might be unsuitable for energetic predictions in A-Mo-O chemical spaces. Our results suggest that the common practice of applying DFT+U to convex hull constructions, especially with high U$_{\text{eff}}$ values derived from fittings, should be carefully evaluated to ensure that ground states are correctly reproduced, with careful consideration of dynamic stability and possible energetically favorable distortions.
\end{abstract}

\maketitle


\section{\label{sec:intro}Introduction}

Modern computational materials science heavily relies on Density Functional Theory (DFT) calculations \cite{PhysRev.136.B864, PhysRev.140.A1133, marzari2021electronic}. A common paradigm in the field is to first construct the energy convex hull from high-throughput (HT) DFT databases, such as the Open Quantum Materials Database (OQMD) \cite{saal2013materials, kirklin2015open}, Materials Project (MP) \cite{jain2013commentary}, and Automatic FLOW (AFLOW) \cite{curtarolo2012aflow}. These convex hulls serve as theoretical references for ground-state phase diagrams and form the basis for various practical tasks, including materials discovery \cite{montoya2020autonomous, ye2022novel, merchant2023scaling}, synthesis prediction \cite{aykol2018thermodynamic, montoya2024ai}, determination of electrochemical stability \cite{singh2017electrochemical}, availability of disordered states \cite{lun2021cation, divilov2024disordered}, and of redox behavior and reaction chemistry \cite{ong2008li, kirklin2013high, aykol2014local}. However, within this formalism, inaccuracies in the DFT-identified ground states and the energy ranking of different phases in any given chemical system will affect all of the aforementioned derived quantities. Transition-metal (TM) oxide systems are particularly challenging for DFT due to varying degrees of electron (de)localization in compounds. Nonetheless, TM oxides are technologically crucial in many application areas (e.g., as energy materials in efforts toward carbon neutrality), and accurate theoretical predictions are essential for accelerating their development.

Unlike in the exact density functional or in the Hartree-Fock (HF) equations, the spurious self-interaction in the Hartree term of the Kohn-Sham equations is not fully cancelled by the exchange term in conventional exchange-correlation functionals based on the Local Density Approximation (LDA) and the Generalized Gradient Approximation (GGA) \cite{perdew1981self}. As approximations that originated from the homogeneous electron gas, these functionals generally correct better for delocalized states but perform poorly on localized states that exhibit stronger self-interaction, leading to overdelocalization. This self-interaction error (SIE) becomes particularly evident in quantities involving comparisons between vastly different localized and delocalized electron states, a common scenario in TM oxides \cite{zhou2004first}. 

To describe localized states better, on-site Hubbard U corrections on DFT (DFT+U) \cite{anisimov1997first, liechtenstein1995density} has become a popular and cost-effective approach that reduces SIE \cite{himmetoglu2014hubbard}. The aforementioned HTDFT databases \cite{kirklin2015open, wang2006oxidation, jain2011formation, calderon2015aflow} typically employ the simplified rotationally invariant approach by Dudarev et al. \cite{ dudarev1998electron}, which relies on only one effective Hubbard U parameter U$_{\text{eff}}$ = U - J, where U and J represent the on-site Coulomb and exchange interactions, respectively. Despite the existence of non-empirical approaches for determining the U value, such as the constrained random phase approximation \cite{vaugier2012hubbard} and linear response approach \cite{cococcioni2005linear}, the U value is commonly employed as an empirical parameter to fit experimentally measured quantities, including band gaps, lattice parameters, redox energies, and formation enthalpies \cite{wang2006oxidation, jain2011formation, aykol2014local, lutfalla2011calibration, capdevila2016performance}. However, the limited transferability of the U value across various aspects has been reported. For instance, the U value that is fit to produce accurate redox reaction energies depends not only on the TM species but also on the oxidation state, local environment, and coordinating ligand~\cite{aykol2014local, capdevila2016performance}. Additionally, suitable U values can vary even when fitted to the same compounds but for different properties \cite{capdevila2016performance}. There is no assurance that a single U value, either fitted for one property or non-empirically derived, will accurately predict all properties, including the correct 0~K ground state and subsequent stability predictions. Despite these challenges, HTDFT databases typically assign a single U value for each TM across chemical spaces to facilitate manageable energy comparisons. Still, these limitations raise concerns about the reliability of DFT convex hulls as demonstrated by Mo-containing oxides in this study, as their conclusions can vary based on this one adjustable parameter.

In this work, we investigate the impact of the U$_{\text{eff}}$ value on layered AMoO$_2$ (A= Li, Na, K) and rutile MoO$_2$, as well as their distorted versions, discovering surprising and drastic reversals in distortion stability (E$_{distorted}$ - E$_{undistorted}$) and corresponding convex hull changes as a function of U$_{\text{eff}}$ value. We perform PBE+U$_{\text{eff}}$ calculations with U$_{\text{eff}}$ value ranging from 0 to 5 eV in steps of 0.5 eV applied to Mo, covering different U values reported in HTDFT dabases and the literature: MP uses U$_{\text{eff}}$ = 4.38 eV by fitting the oxidation reaction energy of Mo$^{4+\to6+}$ \cite{materials_project_hubbard_u}, AFLOW uses U$_{\text{eff}}$ = 2.4 eV, the value implemented in their work of HT electronic band structures \cite{setyawan2011high}, and OQMD does not add U to Mo. Jain et al. \cite{jain2011formation} reported the value as 3.5 eV from the fitting of formation energies for multiple Mo-containing oxides, and the recent work by Moore et al. \cite{moore2024high} reported the mean value as 1.911$\pm$0.318 eV using non-empirical HT linear response. Our work primarily focuses on the technologically important layered LiMoO$_2$, comparing PBE+U$_{\text{eff}}$ results with +U$_{\text{eff}}$ on PBE-D3 van der Waals (vdW) dispersion-correction \cite{grimme2010consistent, aykol2015van} and r$^2$SCAN functionals \cite{furness2020accurate}. We also validated our findings with Heyd-Scuseria-Ernzerhof (HSE06) hybrid functionals \cite{heyd2003hybrid, heyd2004efficient, krukau2006influence, chevrier2010hybrid, seo2015calibrating}, which partially correct SIE by including HF exchange, albeit at considerable computational expense. By carefully examining the stability, bond length, volume, and magnetism in both experimental data and DFT results across various functionals and corrections, along with inspecting the dynamic stability of structures at both high and low U$_{\text{eff}}$, we demonstrate that the anomaly occurs uniquely with a high U$_{\text{eff}}$ correction. For Mo-containing oxides, calculations with low U$_{\text{eff}}$ provide better agreement with experimental observations across AMoO$_2$ for all the aforementioned examined quantities, as far as experimental data is available. Finally, we discuss the uncertainty propagation in HT DFT+U$_{\text{eff}}$ scheme arising from the common practice of empirically fitting and then applying a single constant U$_{\text{eff}}$ across the phase space. This study provides a pathway to expose anomalies induced by high U$_{\text{eff}}$ corrections and the resulting inaccuracies, offering practical approaches to verify physically robust DFT predictions for TM oxides.

\section{\label{sec:method}Methods}
All spin-polarized DFT calculations were performed using the Vienna Ab initio Simulation Package (\texttt{VASP}) \cite{kresse1993ab, kresse1996efficiency, kresse1996efficient} with Projector Augmented Wave (PAW) pseudopotentials \cite{blochl1994projector, kresse1999ultrasoft}. \texttt{VASP} Mo\_sv ($4s^24p^64d^55s^1$) pseudopotentials were used for calculations presented in this letter, while Mo\_pv ($4p^64d^55s^1$) and Mo ($4d^55s^1$) pseudopotentials were checked and also exhibit analogous high U$_{\text{eff}}$ anomaly. Hubbard U corrections were applied using the simplified rotationally invariant scheme proposed by Dudarev et al. \cite{dudarev1998electron}. The vdW-dispersion energy corrections on PBE were introduced using the DFT-D3 method by Grimme et al. \cite{grimme2010consistent} with both zero-damping (PBE+D3) and Becke-Johnson (PBE+D3/BJ) damping variants \cite{grimme2011effect}. Hybrid functional calculations were carried out using HSE06 \cite{heyd2003hybrid, heyd2004efficient, krukau2006influence} with default 25\% mixing of short-range HF exchange. The convergence criteria for structural relaxations were set at 0.01 eV/$\text{\AA}$ and 10$^{-6}$ eV for ionic and electronic loops, respectively. Reported quantities were obtained from additional static calculations with an electronic energy convergence of 10$^{-6}$ eV after the structural relaxations. We employ Gaussian smearing with a 0.05 eV smearing width for structural relaxations and the tetrahedron method with Bl\"{o}ch corrections \cite{blochl1994improved} for static calculations. A plane-wave basis set with an energy cutoff of 520 eV for structural relaxations and 680 eV for static runs was used. The gamma-centered k-meshes for the Brillouin zone integration were generated with a KSPACING of 0.2 $\text{\AA}^{-1}$ for structural relaxations and 0.1 $\text{\AA}^{-1}$ for static runs. We note that the initial geometries for structural optimizations are acquired from OQMD. To reduce bias toward U$_{\text{eff}}$ = 0 and minimize errors from volume changes in structural optimizations, relaxations with the above settings are performed twice at each U$_{\text{eff}}$ value before the final static calculations.  

For the phonon calculations, structures were first further relaxed with convergence criteria of $10^{-8}$ eV for energy and $10^{-3}$ eV/$\text{\AA}$ for forces, followed by single-point atomic displacement calculations with a convergence criterion of $10^{-8}$ eV for energy. Both steps were performed with an energy cutoff of 520 eV, a KSPACING of 0.15 $\text{\AA}^{-1}$, and Gaussian smearing with a 0.05 eV smearing width. All phonon calculations were performed using \texttt{phonopy} \cite{phonopy-phono3py-JPCM, phonopy-phono3py-JPSJ} with atomic displacements of 0.01 $\text{\AA}$, and plots were generated with \texttt{ThermoParser} \cite{spooner2024thermoparser}. For LiMoO$_2$, we used supercell sizes of 4x4x1 of the conventional cell for $R\overline{3}m$ (192 atoms), 1x4x3 of the conventional cell for $C2/m$ (192 atoms), and 2x2x2 of the primitive cell for $P\overline{1}$ (128 atoms). For MoO$_2$, we used supercell sizes of 3x3x4 of the primitive cell for $P4_2/mnm$ (216 atoms) and 2x3x2 of the primitive cell for $P2_1/c$ (144 atoms).

The Special Quasirandom Structure (SQS) method \cite{zunger1990special} was used to generate a 32-atom supercell representing the cation-disordered state of LiMoO$_2$ by employing the integrated cluster expansion toolkit (\texttt{ICET}) \cite{aangqvist2019icet} package and cross-checking with the Alloy Theoretic Automated Toolkit (\texttt{ATAT}) \cite{van2009multicomponent, van2013efficient}.

\section{\label{sec:level1}Results}

\begin{figure} [h!]
\includegraphics[width=\columnwidth]{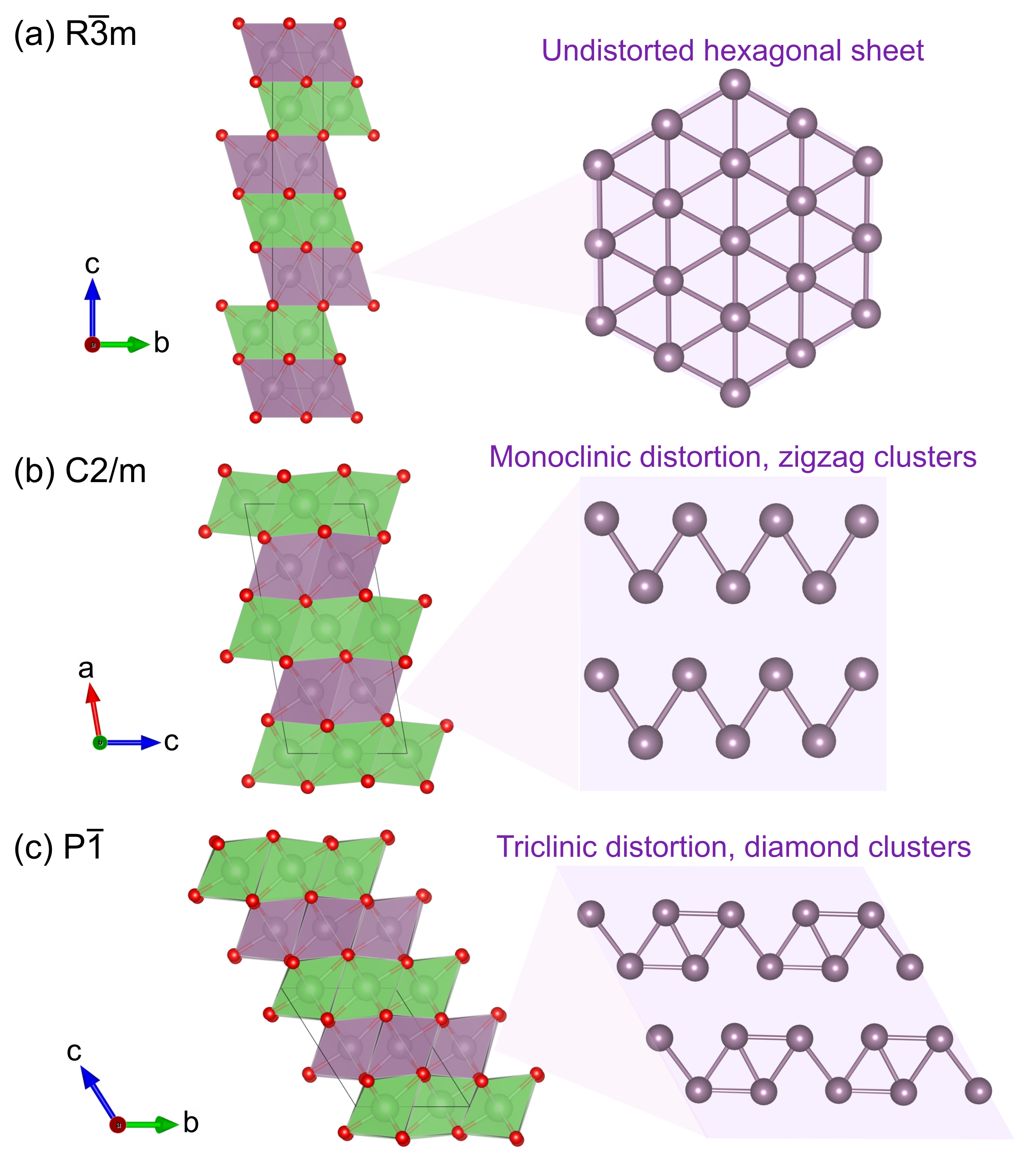}
\caption{\label{fig:structure} Crystal structure prototypes and their corresponding molybdenum layer arrangements (top view) in layered AMoO$_2$ include: (a) an undistorted rhombohedral structure (space group: $R\overline{3}m$), (b) a monoclinic distorted structure ($C2/m$) with zigzag molybdenum clusters, and (c) a triclinic distorted structure ($P\overline{1}$) with diamond molybdenum clusters. \cite{hibble1997true}}
\end{figure}

In Fig. \ref{fig:structure}, we show three crystal structures relevant to the AMoO$_2$ compounds, specifically the layered $R\overline{3}m$ structure and two distorted variants $C2/m$ and $P\overline{1}$.  The structure with the highest symmetry among the three is the rhombohedral $R\overline{3}m$ cell, which is isostructural with the well-known battery cathode material LiCoO$_2$, in which Li and TM are octahedrally coordinated by oxygen atoms and separated into alternating close-packed hexagonal cation layers (Fig.~\ref{fig:structure}a). Many $d^3$ systems, such as NaMoO$_2$ \cite{vitoux2020namoo2}, LiMoS$_2$ \cite{dungey1998structural, petkov2002structure, rocquefelte2000mo}, ReS$_2$, and TcS$_2$ \cite{wildervanck1971dichalcogenides, lamfers1996crystal, fang1997electronic}, are both experimentally and computationally reported to exhibit a distortion involving cluster formation in the TM planes that lowers the symmetry, a phenomenon attributed to a Peierls instability \cite{peierls1996quantum, fang1997electronic, canadell1989origin, vitoux2020namoo2}. If this phenomenon occurs in the $R\overline{3}m$ layered AMoO$_2$, the resulting distortion leads to a $C2/m$ monoclinic cell with zigzag Mo clusters (Fig.~\ref{fig:structure}b) or a more severely distorted $P\overline{1}$ triclinic cell with diamond Mo clusters (Fig.~\ref{fig:structure}c).

\subsection{\label{sec:AllLiMoO2} LiMoO$_2$}
\subsubsection{PBE+U$_{\text{eff}}$ and the varying convex hull}

\begin{figure}
\includegraphics[width=0.9\columnwidth]{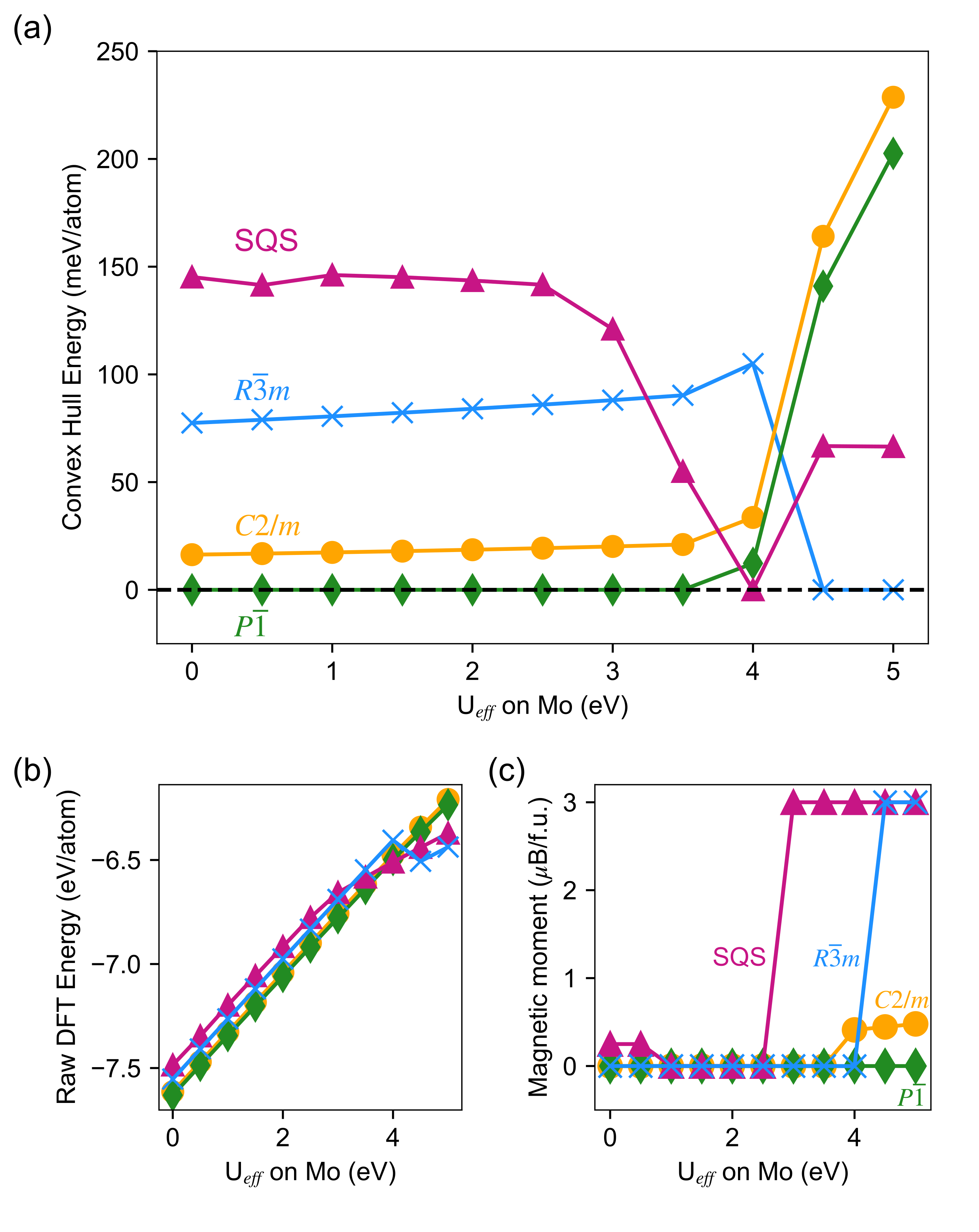}
\caption{\label{fig:LiMoO2} LiMoO$_2$ PBE+U$_{\text{eff}}$ (a) ground states and convex hull energies (E$_{hull}$) that show variations in the most stable state and stability predictions as a function of U$_{\text{eff}}$. Note that the dotted line is simply the convex hull energy (set to 0 by definition). (b) Raw energies used for calculating the convex hull and (c) the corresponding magnetic moment of each prototype and SQS, to demonstrate the correlations between energetic changes and magnetizations as a function of varying U$_{\text{eff}}$.}
\end{figure}

We first examine the ground state and the subsequent structural stability changes corresponding to PBE+U$_{\text{eff}}$ in LiMoO$_2$ in the three crystal structure prototypes of Fig.~\ref{fig:structure}. As shown in Fig.~\ref{fig:LiMoO2}a, the $P\overline{1}$ distorted structure is always on the hull with U$_{\text{eff}}$ $<$ 4 eV on Mo. The convex hull energy (E$_{hull}$) values for three ordered prototypes roughly remain unchanged in this region, with the distorted $C2/m$ slightly above the hull, and the undistorted $R\overline{3}m$ cell exhibiting E$_{hull}$ $>$ 75 meV/atom is the highest energy structure. However, this energy lowering of the distorted structures drastically reverses at U$_{\text{eff}}$ $>$ 4 eV. The undistorted $R\overline{3}m$ cell now falls on the hull, and both $P\overline{1}$ and $C2/m$ distorted structures become energetically unfavorable with E$_{hull}$ over 100 meV/atom, which keeps increasing with a steep slope at higher U$_{\text{eff}}$. Nevertheless, neither distorted structures simply relaxed back into $R\overline{3}m$ from the initial geometry. This topic is discussed further in the following phonon section. The significant reversal is caused by raw energies of $R\overline{3}m$ dropping at U$_{\text{eff}}$ = 4.5 eV, shifting to a line with a smaller slope (see Fig.~\ref{fig:LiMoO2}b) due to more localized on-site occupancies in Mo d-orbitals, as verified in static calculation outputs. The extensive change in orbital occupancies correlates with the emergence of a strongly ferromagnetic (FM) solution with 3 $\mu$B/f.u. in the $R\overline{3}m$ cell (see Fig.~\ref{fig:LiMoO2}c), in which most magnetization happens on Mo. In contrast, the $P\overline{1}$ and $C2/m$ cells show a smooth and slow increase in d-orbital localization, and no sharp modifications in their raw energies or magnetic moments were found. In the OQMD (without U$_{\text{eff}}$ on Mo), the $P\overline{1}$ structure is stable on the convex hull. We note that the competing phases, Li$_2$MoO$_4$ + Mo, are nearly degenerate with the $P\overline{1}$ structure in the OQMD, thus raising the issue of how these competing phases are affected by +U$_{\text{eff}}$ corrections. We examined Li$_2$MoO$_4$ across U$_{\text{eff}}$ from 0 to 5 eV on Mo and found no transitions similar to those in the $R\overline{3}m$ cell. Therefore, we conclude that the Li$_2$MoO$_4$ + Mo competing phases are unlikely to significantly contribute to the stability anomaly in LiMoO$_2$ and are hence not shown in this letter.

\begin{figure*}[ht]
\includegraphics[width=\columnwidth]{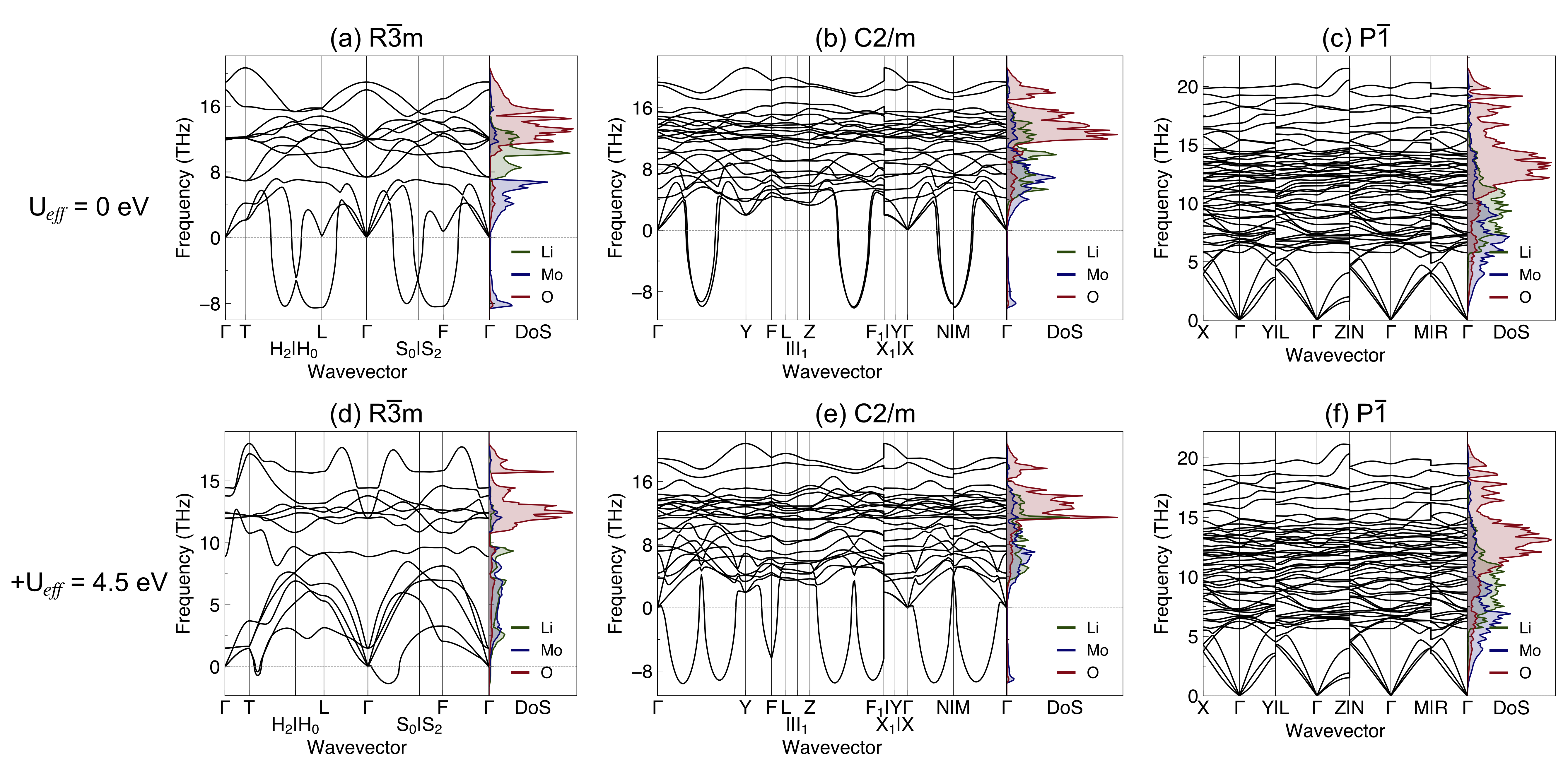}%
\caption{\label{fig:phonon} Phonon band structure and atom-projected density of states of LiMoO$_2$ with PBE+U$_{\text{eff}}$ = 0 and 4.5 eV in the (a)(d) $R\overline{3}m$, (b)(e) $C2/m$, and (c)(f) $P\overline{1}$ structures. The presence of negative phonon modes indicates dynamic instability in the $R\overline{3}m$ and $C2/m$ structures.}
\end{figure*}

Next, we extend our investigation to a rocksalt-based LiMoO$_2$ disordered state, represented by a 32-atom SQS supercell with disordered mixing of Li and Mo on the cation sublattice. We found that SQS not only exhibits profound variations in energies and magnetic moments (though smoother than in the $R\overline{3}m$ cell) but also starts experiencing variations at a lower U$_{\text{eff}}$ = 3 eV than the ordered compound. The SQS energy even reaches the convex hull at U$_{\text{eff}}$ = 4 eV, where it is the only structure in the plot with strong magnetization and d-orbital localization. Notably, the entire SQS E$_{hull}$ curve becomes nonmonotonic and ranges from 150 to 0 meV/atom. Combined with the aforementioned anomaly in distortion stability that already affects the convex hull, this uncertainty in SQS E$_{hull}$ raises concerns in predictions of synthesizability or disordering temperature derived from E$_{hull}$ in TM oxide systems. While accurate simulations of disordered states are crucial for advanced materials discovery, the effect of +U$_{\text{eff}}$ on (dis)ordering is a separate (and more complicated) topic from the distortion stability explored in this work and will be addressed in a subsequent paper. Given these challenges, the remainder of this letter focuses on the anomalies in three layered ordered structures.

We also examined the dynamic stability of the three prototypes by calculating their harmonic phonon dispersions without U$_{\text{eff}}$ and with high U$_{\text{eff}}$ = 4.5 eV, as demonstrated in Fig.~\ref{fig:phonon}. In Fig.~\ref{fig:LiMoO2}a, relaxation to the low-energy geometry does not occur with or without high U$_{\text{eff}}$, but for fundamentally different reasons in each case. With U$_{\text{eff}}$ = 0, the $R\overline{3}m$ and $C2/m$ prototypes exhibit imaginary phonon modes, indicating both structures are dynamically unstable against perturbations. These cells are artificially constrained in the small, high-symmetry primitive cell, and thus are unable to relax to a more energetically-favorable, lower-symmetry $P\overline{1}$ structure. The $P\overline{1}$ cell is dynamically stable, consistent with the finding in Fig.~\ref{fig:LiMoO2}a that this distorted structure is the thermodynamically stable ground state in PBE with U$_{\text{eff}}$ = 0. Next, we applied a high U$_{\text{eff}}$ of 4.5 eV and found that the $R\overline{3}m$ and $C2/m$ cells still exhibit dynamic instability \footnote{There is a possibility these instabilities arise from small numerical errors in the forces or supercell size effects. Some of the observed imaginary phonon modes occur at wavevectors that are incommensurate with the supercell and are thus not sampled exactly. We cannot rule out the possibility that these structures are instead dynamically stable but close to instability. Further confirmation of dynamic (in)stability would require relaxation of long period supercells that are prohibitively expensive to calculate. Nevertheless, small instabilities are unlikely to lead to large changes in energy after relaxation and thus do not significantly change our conclusions}. The $R\overline{3}m$ cell, despite appearing to be the lowest energy structure among the three prototypes when using U$_{\text{eff}}$ = 4.5 eV, has a very weak dynamic instability. We note that anharmonic effects might stabilize structures at finite-temperature that are dynamically unstable at 0 K, which could be especially important when comparing to experimental results
\cite{PhysRevB.92.054301, PhysRevLett.117.075502, PhysRevLett.125.085901}. In contrast, the $P\overline{1}$ structure is dynamically stable even when its E$_{hull}$ is over 100 meV/atom. This observation is consistent with previous structural relaxation results that the $P\overline{1}$ cell did not relax back into low-energy $R\overline{3}m$ at U$_{\text{eff}}$ = 4.5 eV, suggesting a barrier between the $P\overline{1}$ and $R\overline{3}m$ geometries. Although the $C2/m$ cell is dynamically unstable, we assume that energy barriers between the initial $C2/m$ geometry and $R\overline{3}m$ prevent direct relaxation to $R\overline{3}m$.

Up to this point, we have demonstrated that high or low U$_{\text{eff}}$ in PBE can lead to drastically different results, but there is no decisive evidence from theoretical stability alone to determine which choice more accurately describes the stability of crystal structure distortion in LiMoO$_2$. Unfortunately, identifying the experimental ground truth of LiMoO$_2$ is challenging as its measured crystallography remains uncertain; various symmetries---$R\overline{3}m$ \cite{aleandri1988hexagonal, barker2003lithium, barker2003synthesis}, $C2/m$ \cite{hibble1995local, hibble1997true, ben2012structural, ramana2021growth}, or the mixture of both \cite{mikhailova2011layered}, have been reported in the literature. Unlike other $d^3$ systems, no diamond clusters have been experimentally observed in layered LiMoO$_2$, although Hibble and co-workers \cite{hibble1997true} stated their measurements do not rule out the possibility of diamond clusters. Moreover, due to the similar ionic radii of Li$^{1+}$ and Mo$^{3+}$, Li mixing in Mo layers could occur \cite{mikhailova2011layered,  vitoux2020namoo2}, making the finite-temperature measured structures differ from the ordered structures in DFT calculations and thus unsuitable for precise comparisons. As a result, it is difficult to evaluate the stability among the three ordered and stoichiometric LiMoO$_2$ structures solely from experimental data. In the following section, we provide further computational insights by employing advanced functionals and beyond-DFT methods to determine whether high or low U$_{\text{eff}}$ provides more reliable structural relaxations and energetic descriptions.  

\subsubsection{Distortion stability in multiple functionals and corrections}
\begin{figure} 
\includegraphics[width=0.7\columnwidth]{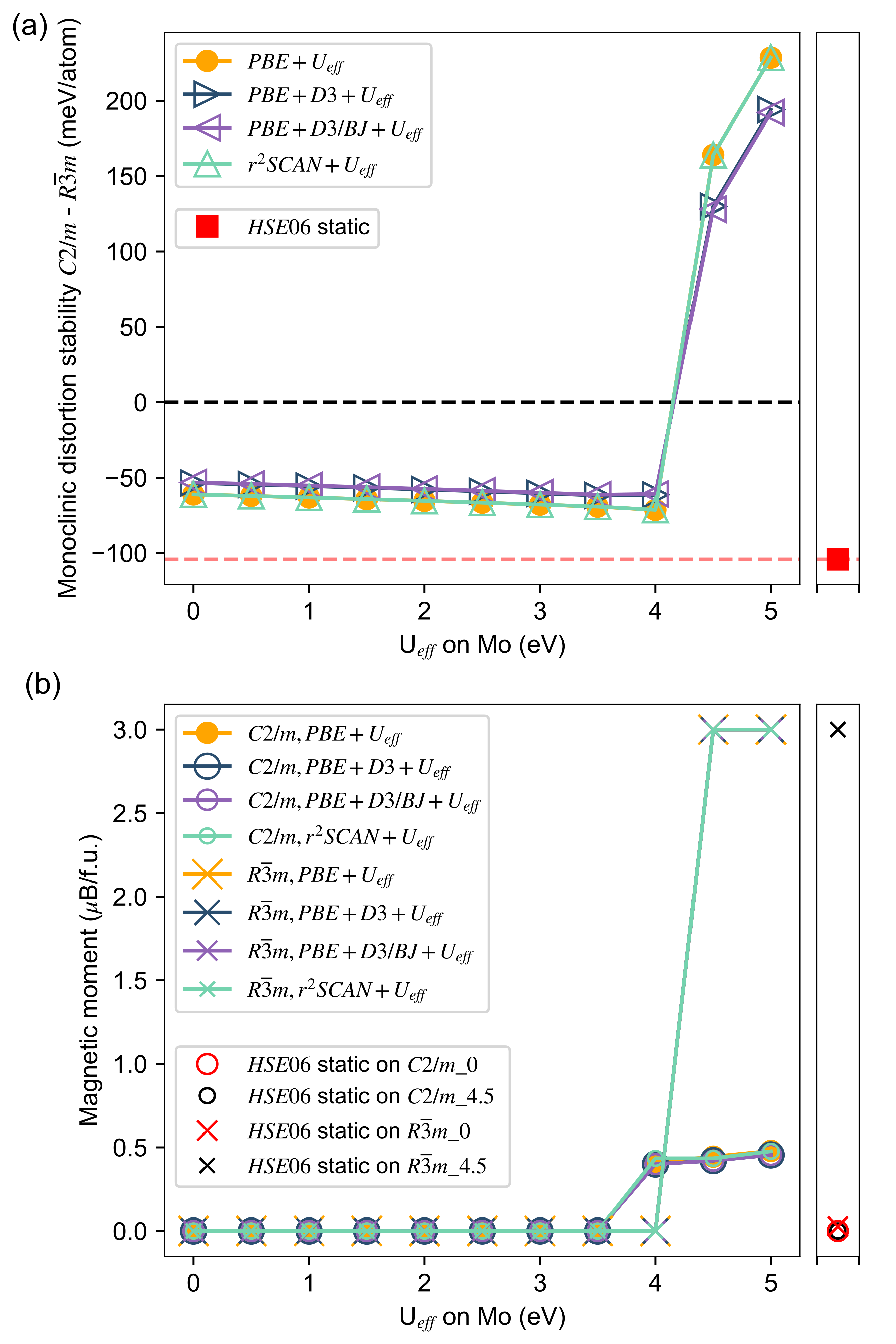}
\caption{\label{fig:HSE} LiMoO$_2$ (a) Monoclinic distortion stability: $C2/m$ - $R\overline{3}m$ and (b) cell magnetic moment, both from various functionals and corrections. HSE06 results are shown in the separate box on the right, with additional details provided in Table~\ref{tab:table1}. The HSE06 static monoclinic distortion stability was calculated using $C2/m\_0$ and $R\overline{3}m\_0$, the lower energy structures within their respective symmetries. Anomalous stability reversal at high U$_{\text{eff}}$ is observed across DFT variants with +U$_{\text{eff}}$, while HSE06 results indicate that distortions should be energetically favorable.}
\end{figure}

 As shown in Fig.~\ref{fig:LiMoO2}, the $P\overline{1}$ and $C2/m$ cells exhibit similar behavior when compared to the $R\overline{3}m$ cell, which allows us to simplify the analysis to a question of whether distortions should be favorable. The $C2/m$ cell is computationally less expensive to calculate than $P\overline{1}$ while showing a similar trend for distortion stability.  Additionally, the $C2/m$ and $R\overline{3}m$ cells are the two debated structures for the synthesized phase of LiMoO$_2$ in the experimental literature. For these reasons, we focus on the monoclinic distortion stability, defined as the energy difference per atom between the $C2/m$ and $R\overline{3}m$ cells, as the primary research target in this section. Previous literature \cite{isaacs2020prediction, chevrier2010hybrid, seo2015calibrating, aykol2015van} has found that the choice of exchange correlation and corrections in DFT can affect the energetics of lithium transition metal oxides, prompting us to test our PBE calculations against vdW-dispersion energy corrections PBE+D3 and PBE+D3/BJ, and r$^2$SCAN functionals, as shown in Fig.~\ref{fig:HSE}. However, the same drastic changes in distortion stability and magnetic moments were consistently observed in all these variants, and thus this anomaly is a general issue of the +U$_{\text{eff}}$ implementation in DFT. 

 To address the SIE in TM oxides without applying +U$_{\text{eff}}$, we performed HSE06 static calculations on structures relaxed by PBE+U$_{\text{eff}}$ = 0 and 4.5 eV (denoted as \_0 and \_4.5 from here on) for both $C2/m$ and $R\overline{3}m$ cells, with results organized in Table~\ref{tab:table1}. While the mixing parameter in HSE06 can still be practically treated as an empirical fitting parameter \cite{seo2015calibrating}, we used the default value of 25\%, derived from perturbation theory \cite{perdew1996rationale}. The data in Table~\ref{tab:table1} demonstrate that using the relaxed PBE+U$_{\text{eff}}$ = 0 geometries gives lower HSE06 energies for both $C2/m$ and $R\overline{3}m$ structures (i.e., $C2/m\_0$ is lower than $C2/m\_4.5$ and $R\overline{3}m\_0$ is lower than $R\overline{3}m\_4.5$). We observed that $R\overline{3}m\_4.5$ is the only one that converges to a strong FM state. However, compared to the converged low magnetic moment state in $R\overline{3}m\_0$, the $R\overline{3}m\_4.5$ cell is unstable by 114 meV/atom. Actually, $C2/m\_0$, $C2/m\_4.5$, and $R\overline{3}m\_0$ all converge to a nonmagnetic (NM) or near-zero magnetic moment state, regardless of the strength of the initial spin settings in our testings. It is also evident that $C2/m$, irrespective of U$_{\text{eff}}$ in structural relaxations, is significantly more stable than $R\overline{3}m$, indicating an energetically favorable distortion tendency as observed in other $d^3$ systems. The HSE06 results are qualitatively consistent with the ``low-U'' results from PBE+U$_{\text{eff}}$, and in conflict with the ``high-U'' results, showing an energetic lowering of the distorted structure. 

\begin{table} [h]
\caption{\label{tab:table1}%
The volume and Mo-Mo bond distance (d(Mo-Mo)) of input structures and HSE06 static calculation results. 
}
\begin{ruledtabular}
\begin{tabular}{c c c c c}
\textrm{Structure} &
\multicolumn{1}{c}{\textrm{Volume}} &
\multicolumn{1}{c}{\textrm{d(Mo-Mo)}} &
\multicolumn{1}{c}{\textrm{Energy relative to $C2/m\_0$}} &
\multicolumn{1}{c}{\textrm{Magmom}} \\
&
\multicolumn{1}{c}{(\text{\AA}$^3$/atom)} &
\multicolumn{1}{c}{(\text{\AA})} &
\multicolumn{1}{c}{(meV/atom)} &
\multicolumn{1}{c}{($\mu_B$/f.u.)} \\
\colrule
$C2/m\_0$ & 9.35 & 2.60, 2.80 & 0 & 0.000 \\
$C2/m\_4.5$ & 9.62 & 2.59, 2.89 & 28 & 0.000 \\
$R\overline{3}m\_0$ & 9.12 & 2.79 & 104 & 0.027 \\
$R\overline{3}m\_4.5$ & 10.94 & 3.20 & 218 & 3.000 \\
\end{tabular}
\end{ruledtabular}
\end{table}
 
 Moreover, the $R\overline{3}m\_4.5$ cell, which converges to an FM solution, exhibits a volume and Mo-Mo bond distance (d(Mo-Mo)) that are more than 10\% larger than those of all other NM cells. For experimentally measured $R\overline{3}m$ compounds, Aleandri and McCarley reported a d(Mo-Mo) of 2.866 \AA\ \cite{aleandri1988hexagonal}, and Mikhailova et al. reported 9.25 \AA$^3$/atom for the volume \cite{mikhailova2011layered}. For experimentally measured $C2/m$ compounds, Hibbel et al. reported d(Mo-Mo) values of 2.56 Å and 2.71 Å for short (nearest neighbors on the chain) and long (same-row neighbors) Mo-Mo bonds, respectively, in their EXAFS study \cite{hibble1995local}, and later reported the short bond distance as 2.618 \AA\ using neutron scattering \cite{hibble1997true}. In fact, we reviewed the lattice parameters reported in the literature for synthesized LiMoO$_2$ \cite{aleandri1988hexagonal, barker2003lithium, barker2003synthesis,hibble1995local, hibble1997true, ben2012structural, mikhailova2011layered}, regardless of their phase classification, and found that all their corresponding volumes approximately range from 9.2 to 9.5 \AA$^3$/atom. The FM $R\overline{3}m\_4.5$ cell, which has a volume of 10.94 \AA$^3$/atom and d(Mo-Mo) = 3.20 \AA, deviates considerably from synthesized compounds, whereas other NM cells maintain volumes and bond distances close to the experimental values. To sum up, the observed high +U$_{\text{eff}}$ irregularities might be artifacts of the DFT+U$_{\text{eff}}$ method, as the FM $R\overline{3}m\_{4.5}$, predicted to be the ground state at high U$_{\text{eff}}$, is highly unstable in HSE06 results, and its structure deviates from the range of all experimental measured volumes and d(Mo-Mo).
 
\subsection{\label{sec:NaKMoO2} NaMoO$_2$ and KMoO$_2$} 
We next extend our investigation to NaMoO$_2$ and KMoO$_2$ with the same layered configurations. While NaMoO$_2$ was reported as $R\overline{3}m$ symmetry in a study dating back to 1969 \cite{ringenbach1969hatterer}, recent work by Vitoux et al. \cite{vitoux2020namoo2} reported a highly crystalline NaMoO$_2$ in a $P\overline{1}$ cell, observing diamond clusters. Moreover, they performed PBE without U and found the triclinic distortion to be stable, with the energy difference of -150 meV/atom ($P\overline{1}$ - $R\overline{3}m$) and the $P\overline{1}$ cell being NM and semiconducting, consistent with their experimental measurements. They also reported that a +U correction leads to the disappearance of Mo clusters. Since $C2/m$ is not reported in either NaMoO$_2$ or KMoO$_2$, we shifted our research focus from the monoclinic to the triclinic distortion stability in this section. We note that in OQMD, the $P\overline{1}$ structures of NaMoO$_2$ (KMoO$_2$) are 64 meV/atom (266 meV/atom) above the ground state of Na$_2$MoO$_4$ (K$_2$MoO$_4$) + Mo. Therefore, the discussion here pertains only to distortion stability and is not relevant to the convex hull.

\begin{figure} [h!]
\includegraphics[width=\columnwidth]{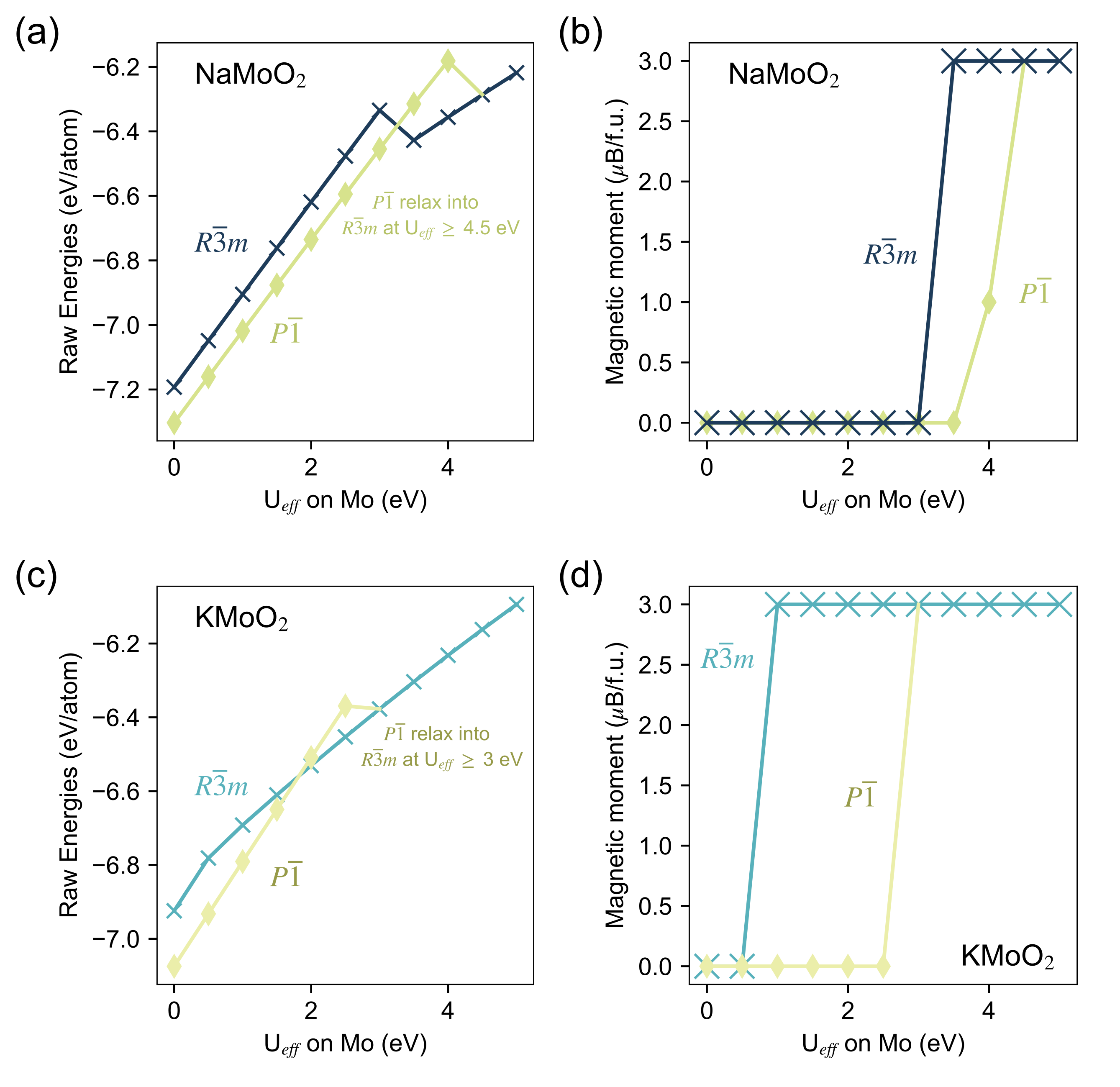}
\caption{\label{fig:NaKMoO2data} PBE+U$_{\text{eff}}$ results for $R\overline{3}m$ and $P\overline{1}$ NaMoO$_2$ (a) raw energies, (b) magnetic moments, and KMoO$_2$ (c) raw energies, (d) magnetic moments, as functions of U$_{\text{eff}}$.}
\end{figure}

As shown in Fig.~\ref{fig:NaKMoO2data}, we reproduced the reversal of triclinic distortion stability in NaMoO$_2$ and KMoO$_2$; however, these compositions exhibited different behaviors compared to the reversal observed in LiMoO$_2$. In NaMoO$_2$, the downshift of energy of the $R\overline{3}m$ cell occurs earlier at U$_{\text{eff}}$ = 3.5 eV. The $P\overline{1}$ cell becomes magnetic at U$_{\text{eff}}$ $\geq$ 4 eV, with its magnetization reaching 3 $\mu$B/f.u at U$_{\text{eff}}$ = 4.5 eV, and the cell relaxes into the higher symmetry $R\overline{3}m$, which corresponds to the disappearance of Mo clusters. In KMoO$_2$, while the reversal persists, the $R\overline{3}m$ cell transitions to the FM state at U$_{\text{eff}}$ $\geq$ 1 eV. The $P\overline{1}$ cell transitions to the FM state at U$_{\text{eff}}$ $\geq$ 3 eV, relaxing into the $R\overline{3}m$ structure. Considering the structural and chemical similarity of the $R\overline{3}m$ and $P\overline{1}$ structures across three AMoO$_2$ systems, the surprising differences in their responses to varying U$_{\text{eff}}$ values again demonstrate the uncertainty of the DFT+U$_{\text{eff}}$ approach in Mo oxide systems. 

\subsection{\label{sec:MoO2} MoO$_2$} 

\begin{figure} [!b]
\includegraphics[width=\columnwidth]{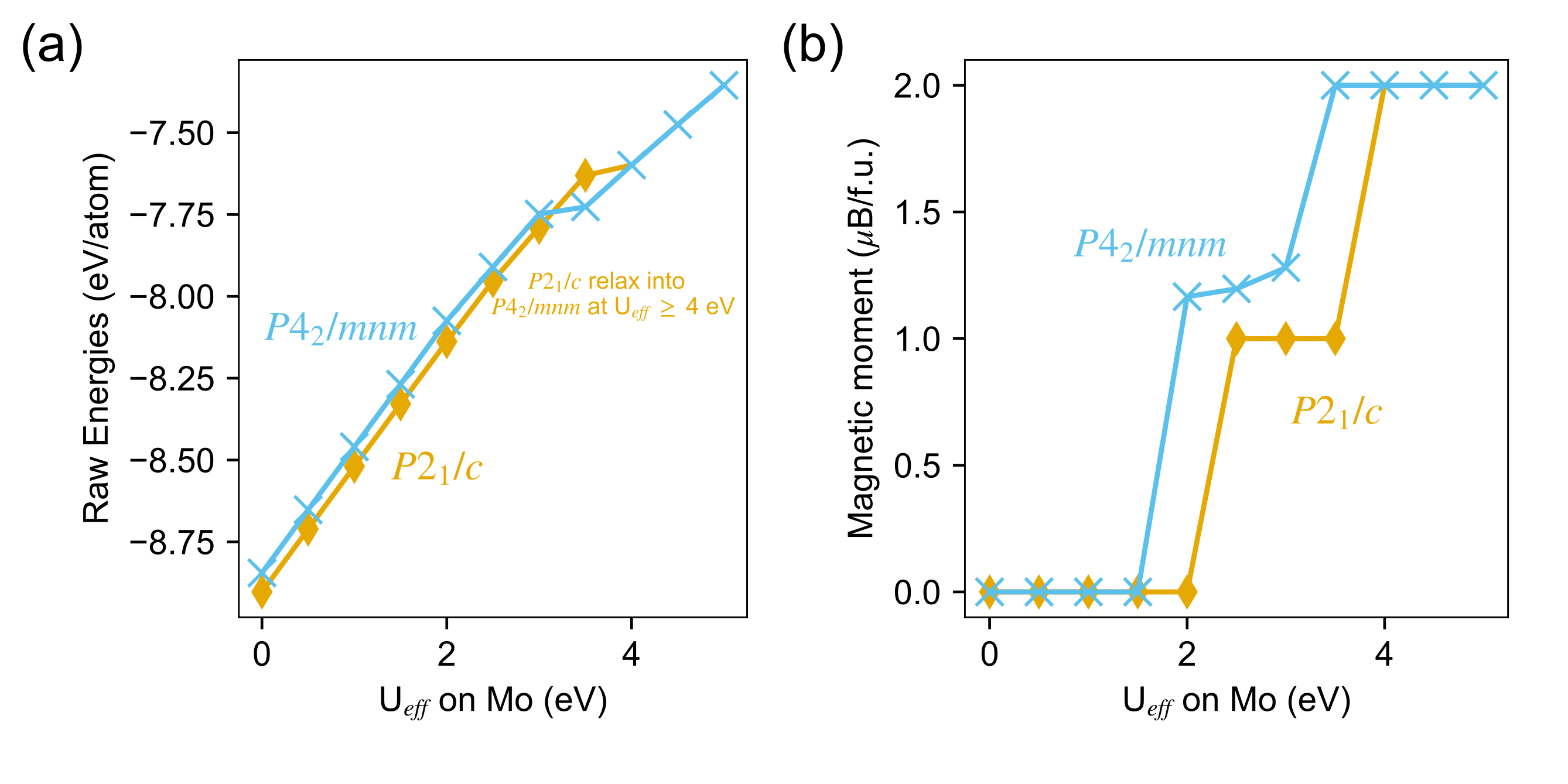}
\caption{\label{fig:MoO2data} PBE+U$_{\text{eff}}$ results for $P4_2/mnm$ and $P2_1/c$ MoO$_2$ (a) raw energies, and (b) magnetic moments as functions of U$_{\text{eff}}$.}
\end{figure}

\begin{figure*}[!t]
\includegraphics[width=1\columnwidth]{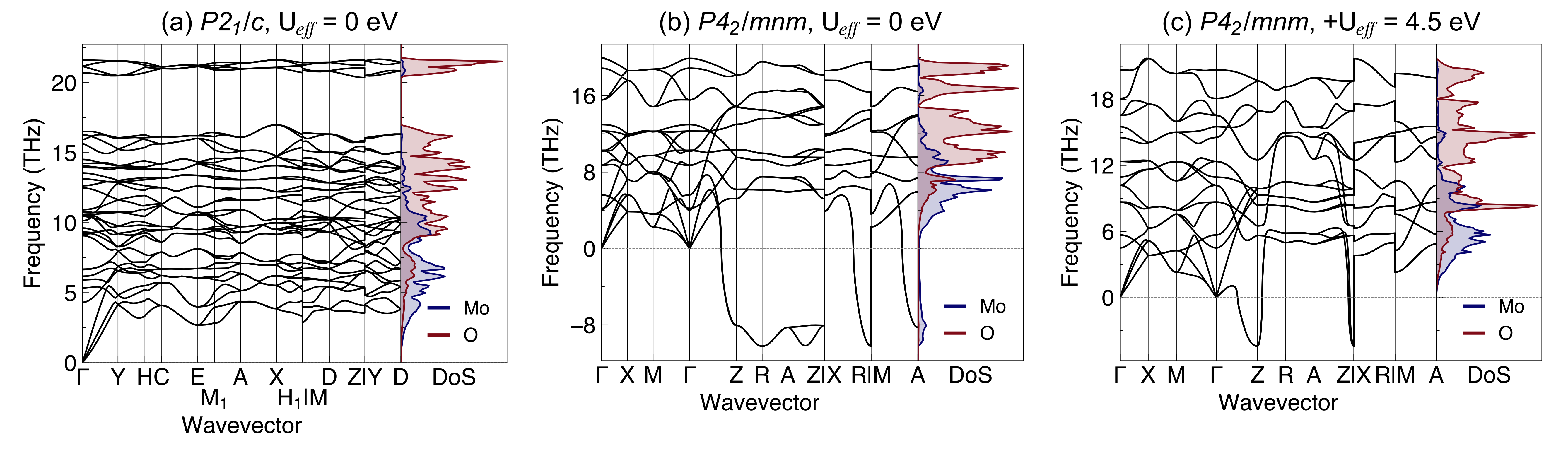} 
\caption{\label{fig:MoO2phonon}  MoO$_2$ phonon band structure and atom-projected density of states for (a) $P2_1/c$ with PBE U$_{\text{eff}}=0$ and $P4_2/mnm$ with PBE (b) U$_{\text{eff}}=0$ (c) +U$_{\text{eff}}$ = 4.5 eV. The $P2_1/c$ cell at U$_{\text{eff}}$ = 4.5 eV relaxes into $P4_2/mnm$. }
\end{figure*} 

Finally, we expanded our study to include rutile-like MoO$_2$, which has a different structure and molybdenum oxidation state, and is expected to behave differently with +U$_{\text{eff}}$. Nevertheless, MoO$_2$ exhibits similar distorted vs. undistorted structural competition on the convex hull. As shown in  Fig.~\ref{fig:MoO2data}, we again observed the distortion stability reversal in distorted vs. high-symmetry rutile structures ($P2_1/c$ vs. $P4_2/mnm$) in MoO$_2$. It is well-known that MoO$_2$ exhibits the $P2_1/c$ structure as its stable synthesized phase \cite{brandt1967refinement, rogers1969crystal, cox1982neutron, bolzan1995neutron, leisegang2005situ}, which is correctly predicted with PBE at low U$_{\text{eff}}$. However, the energy anomaly again occurs at U$_{\text{eff}}$ = 3.5 eV, and the $P2_1/c$ cell relaxes into the higher symmetry $P4_2/mnm$ cell at U$_{\text{eff}}$ $\geq$ 4 eV. The MoO$_2$ magnetization and the corresponding d-orbital localization, again observed in static calculation outputs, as a function of U$_{\text{eff}}$, behave differently compared to layered AMoO$_2$. Specifically, this $d^2$ system exhibits a two-step pattern, illustrating the diverse reactions of molybdenum oxides to changes in U$_{\text{eff}}$. We also examine the dynamic stability of the $P2_1/c$ and $P4_2/mnm$ cells, as shown in Fig.~\ref{fig:MoO2phonon}. The $P2_1/c$  distorted rutile structure is dynamically stable without U$_{\text{eff}}$, while the $P4_2/mnm$ cell exhibits dynamical instability both without U$_{\text{eff}}$ and with high U$_{\text{eff}}$. This behavior is similar to what we have observed in $R\overline{3}m$ LiMoO$_2$, suggesting the energy can be further lowered through relaxation along imaginary modes.

Since MoO$_2$ is commonly included in the U$_{\text{eff}}$ value fitting dataset for molybdenum oxide systems, accurately describing MoO$_2$ in DFT is fundamental. The best mean absolute errors reported in U$_{\text{eff}}$ fitting literature are on the order of 10 to 100 meV/atom \cite{jain2011formation, aykol2014local}, while the energy difference between the $P2_1/c$ and $P4_2/mnm$ cells ranges from -58 to 96 meV/atom for U$_{\text{eff}}$ = 0 and 3.5 eV. Incorporating slight distortions to the rutile structure can easily be impactful for fitting MoO$_2$. Therefore, a careful selection of the initial input structure in DFT, matching the actual synthesized structures responsible for fitted experimental quantities, is essential. The potential uncertainty in U$_{\text{eff}}$ fittings suggests the need to disclose DFT relaxed structures used in U$_{\text{eff}}$ fittings.

The uncertainty in U$_{\text{eff}}$ fittings propagates, leading to inaccuracies in the predicted properties of Mo oxides across all compositions in their extended phase spaces when a single U$_{\text{eff}}$ value is applied. This impact is further amplified by the sensitivity of properties to  U$_{\text{eff}}$, as evidenced by multiple anomalous stability reversals in AMoO$_2$ and the extreme case of SQS LiMoO$_2$, where U$_{\text{eff}}$ between 3 to 5 eV can cause nonmonotonic variations of 150 meV/atom in its E$_{hull}$. This issue from DFT+U$_{\text{eff}}$ implementation is ubiquitous across functionals and can potentially be generalized to other TM phase spaces.

Correct DFT energetic predictions are foundational to the enterprise of computational materials discovery and calculations of their extended functional properties. When using DFT+U$_{\text{eff}}$, especially with high U$_{\text{eff}}$ values for convex hull constructions and stability predictions, it is essential to carefully assess calculations to ensure accurate ground states and replicate experimental observables. Practical biases, such as applying a constant U$_{\text{eff}}$ without accounting for limited transferability and underestimating the sensitivity of target properties to the U$_{\text{eff}}$ value may lead to overlooked dynamic instability and/or energetically favorable distortions, potentially introducing uncertainties up to the order of 100 meV/atom in energetic predictions. This study calls for a comprehensive re-examination of DFT+U$_{\text{eff}}$ in TM oxides, aiming to pave the way for more effective and accurate materials design campaigns.

\section*{Data Availability Statement}
Structures studied in this work and their standard first-principles calculations (PBE) will be released in the Open Quantum Materials Database after publication. The remaining data from advanced calculations used to produce this manuscript is available upon reasonable request.  

\begin{acknowledgments}
T.L. and A.S.-C. acknowledge funding from the Toyota Research Institute (overall leadership of the project, DFT calculations). D.G. acknowledges support from the US Department of Commerce and National Institute of Standards and Technology as part of the Center for Hierarchical Materials Design (CHiMaD) under award no. 70NANB19H005 (phonon calculations). T.L. acknowledges the support from Taiwanese Government Fellowship for Overseas Study. H.K. acknowledges the financial support from the Office of Naval Research, under Award N00014-23-1-2311. A.S.-C. acknowledges the financial support to the National Agency for Research and Development (ANID)/DOCTORADO BECAS CHILE/2018 - 56180024. The authors acknowledge computational resources from the Quest high-performance computing facility at Northwestern University, which is jointly supported by the Office of the Provost, the Office for Research, and Northwestern University Information Technology (DFT calculations). Phonon calculations were performed using resources at the National Energy Research Scientific Computing Center (NERSC), a Department of Energy Office of Science User Facility, under NERSC award BES-ERCAP0027615.
\end{acknowledgments}

\section*{Author Contributions}
The manuscript was written by T.L. with help from other authors. T.L. performed DFT calculations and data analysis with help from H.K. and A.S.-C. D.G. performed phonon calculations. C.W. and S.B.T. supervised the project. All authors contributed to discussions.

\bibliography{apssamp}

\begin{thebibliography}{85}%
\makeatletter
\providecommand \@ifxundefined [1]{%
 \@ifx{#1\undefined}
}%
\providecommand \@ifnum [1]{%
 \ifnum #1\expandafter \@firstoftwo
 \else \expandafter \@secondoftwo
 \fi
}%
\providecommand \@ifx [1]{%
 \ifx #1\expandafter \@firstoftwo
 \else \expandafter \@secondoftwo
 \fi
}%
\providecommand \natexlab [1]{#1}%
\providecommand \enquote  [1]{``#1''}%
\providecommand \bibnamefont  [1]{#1}%
\providecommand \bibfnamefont [1]{#1}%
\providecommand \citenamefont [1]{#1}%
\providecommand \href@noop [0]{\@secondoftwo}%
\providecommand \href [0]{\begingroup \@sanitize@url \@href}%
\providecommand \@href[1]{\@@startlink{#1}\@@href}%
\providecommand \@@href[1]{\endgroup#1\@@endlink}%
\providecommand \@sanitize@url [0]{\catcode `\\12\catcode `\$12\catcode `\&12\catcode `\#12\catcode `\^12\catcode `\_12\catcode `\%12\relax}%
\providecommand \@@startlink[1]{}%
\providecommand \@@endlink[0]{}%
\providecommand \url  [0]{\begingroup\@sanitize@url \@url }%
\providecommand \@url [1]{\endgroup\@href {#1}{\urlprefix }}%
\providecommand \urlprefix  [0]{URL }%
\providecommand \Eprint [0]{\href }%
\providecommand \doibase [0]{https://doi.org/}%
\providecommand \selectlanguage [0]{\@gobble}%
\providecommand \bibinfo  [0]{\@secondoftwo}%
\providecommand \bibfield  [0]{\@secondoftwo}%
\providecommand \translation [1]{[#1]}%
\providecommand \BibitemOpen [0]{}%
\providecommand \bibitemStop [0]{}%
\providecommand \bibitemNoStop [0]{.\EOS\space}%
\providecommand \EOS [0]{\spacefactor3000\relax}%
\providecommand \BibitemShut  [1]{\csname bibitem#1\endcsname}%
\let\auto@bib@innerbib\@empty
\bibitem [{\citenamefont {Hohenberg}\ and\ \citenamefont {Kohn}(1964)}]{PhysRev.136.B864}%
  \BibitemOpen
  \bibfield  {author} {\bibinfo {author} {\bibfnamefont {P.}~\bibnamefont {Hohenberg}}\ and\ \bibinfo {author} {\bibfnamefont {W.}~\bibnamefont {Kohn}},\ }\bibfield  {title} {\bibinfo {title} {{Inhomogeneous Electron Gas}},\ }\href {https://doi.org/10.1103/PhysRev.136.B864} {\bibfield  {journal} {\bibinfo  {journal} {Phys. Rev.}\ }\textbf {\bibinfo {volume} {136}},\ \bibinfo {pages} {B864} (\bibinfo {year} {1964})}\BibitemShut {NoStop}%
\bibitem [{\citenamefont {Kohn}\ and\ \citenamefont {Sham}(1965)}]{PhysRev.140.A1133}%
  \BibitemOpen
  \bibfield  {author} {\bibinfo {author} {\bibfnamefont {W.}~\bibnamefont {Kohn}}\ and\ \bibinfo {author} {\bibfnamefont {L.~J.}\ \bibnamefont {Sham}},\ }\bibfield  {title} {\bibinfo {title} {{Self-Consistent Equations Including Exchange and Correlation Effects}},\ }\href {https://doi.org/10.1103/PhysRev.140.A1133} {\bibfield  {journal} {\bibinfo  {journal} {Phys. Rev.}\ }\textbf {\bibinfo {volume} {140}},\ \bibinfo {pages} {A1133} (\bibinfo {year} {1965})}\BibitemShut {NoStop}%
\bibitem [{\citenamefont {Marzari}\ \emph {et~al.}(2021)\citenamefont {Marzari}, \citenamefont {Ferretti},\ and\ \citenamefont {Wolverton}}]{marzari2021electronic}%
  \BibitemOpen
  \bibfield  {author} {\bibinfo {author} {\bibfnamefont {N.}~\bibnamefont {Marzari}}, \bibinfo {author} {\bibfnamefont {A.}~\bibnamefont {Ferretti}},\ and\ \bibinfo {author} {\bibfnamefont {C.}~\bibnamefont {Wolverton}},\ }\bibfield  {title} {\bibinfo {title} {Electronic-structure methods for materials design},\ }\href@noop {} {\bibfield  {journal} {\bibinfo  {journal} {Nat. Mater.}\ }\textbf {\bibinfo {volume} {20}},\ \bibinfo {pages} {736} (\bibinfo {year} {2021})}\BibitemShut {NoStop}%
\bibitem [{\citenamefont {Saal}\ \emph {et~al.}(2013)\citenamefont {Saal}, \citenamefont {Kirklin}, \citenamefont {Aykol}, \citenamefont {Meredig},\ and\ \citenamefont {Wolverton}}]{saal2013materials}%
  \BibitemOpen
  \bibfield  {author} {\bibinfo {author} {\bibfnamefont {J.~E.}\ \bibnamefont {Saal}}, \bibinfo {author} {\bibfnamefont {S.}~\bibnamefont {Kirklin}}, \bibinfo {author} {\bibfnamefont {M.}~\bibnamefont {Aykol}}, \bibinfo {author} {\bibfnamefont {B.}~\bibnamefont {Meredig}},\ and\ \bibinfo {author} {\bibfnamefont {C.}~\bibnamefont {Wolverton}},\ }\bibfield  {title} {\bibinfo {title} {{Materials design and discovery with high-throughput density functional theory: the open quantum materials database (OQMD)}},\ }\href@noop {} {\bibfield  {journal} {\bibinfo  {journal} {JOM}\ }\textbf {\bibinfo {volume} {65}},\ \bibinfo {pages} {1501} (\bibinfo {year} {2013})}\BibitemShut {NoStop}%
\bibitem [{\citenamefont {Kirklin}\ \emph {et~al.}(2015)\citenamefont {Kirklin}, \citenamefont {Saal}, \citenamefont {Meredig}, \citenamefont {Thompson}, \citenamefont {Doak}, \citenamefont {Aykol}, \citenamefont {R{\"u}hl},\ and\ \citenamefont {Wolverton}}]{kirklin2015open}%
  \BibitemOpen
  \bibfield  {author} {\bibinfo {author} {\bibfnamefont {S.}~\bibnamefont {Kirklin}}, \bibinfo {author} {\bibfnamefont {J.~E.}\ \bibnamefont {Saal}}, \bibinfo {author} {\bibfnamefont {B.}~\bibnamefont {Meredig}}, \bibinfo {author} {\bibfnamefont {A.}~\bibnamefont {Thompson}}, \bibinfo {author} {\bibfnamefont {J.~W.}\ \bibnamefont {Doak}}, \bibinfo {author} {\bibfnamefont {M.}~\bibnamefont {Aykol}}, \bibinfo {author} {\bibfnamefont {S.}~\bibnamefont {R{\"u}hl}},\ and\ \bibinfo {author} {\bibfnamefont {C.}~\bibnamefont {Wolverton}},\ }\bibfield  {title} {\bibinfo {title} {{The Open Quantum Materials Database (OQMD): assessing the accuracy of DFT formation energies}},\ }\href@noop {} {\bibfield  {journal} {\bibinfo  {journal} {npj Comput. Mater.}\ }\textbf {\bibinfo {volume} {1}},\ \bibinfo {pages} {1} (\bibinfo {year} {2015})}\BibitemShut {NoStop}%
\bibitem [{\citenamefont {Jain}\ \emph {et~al.}(2013)\citenamefont {Jain}, \citenamefont {Ong}, \citenamefont {Hautier}, \citenamefont {Chen}, \citenamefont {Richards}, \citenamefont {Dacek}, \citenamefont {Cholia}, \citenamefont {Gunter}, \citenamefont {Skinner}, \citenamefont {Ceder} \emph {et~al.}}]{jain2013commentary}%
  \BibitemOpen
  \bibfield  {author} {\bibinfo {author} {\bibfnamefont {A.}~\bibnamefont {Jain}}, \bibinfo {author} {\bibfnamefont {S.~P.}\ \bibnamefont {Ong}}, \bibinfo {author} {\bibfnamefont {G.}~\bibnamefont {Hautier}}, \bibinfo {author} {\bibfnamefont {W.}~\bibnamefont {Chen}}, \bibinfo {author} {\bibfnamefont {W.~D.}\ \bibnamefont {Richards}}, \bibinfo {author} {\bibfnamefont {S.}~\bibnamefont {Dacek}}, \bibinfo {author} {\bibfnamefont {S.}~\bibnamefont {Cholia}}, \bibinfo {author} {\bibfnamefont {D.}~\bibnamefont {Gunter}}, \bibinfo {author} {\bibfnamefont {D.}~\bibnamefont {Skinner}}, \bibinfo {author} {\bibfnamefont {G.}~\bibnamefont {Ceder}}, \emph {et~al.},\ }\bibfield  {title} {\bibinfo {title} {{Commentary: The Materials Project: A materials genome approach to accelerating materials innovation}},\ }\href@noop {} {\bibfield  {journal} {\bibinfo  {journal} {APL Mater.}\ }\textbf {\bibinfo {volume} {1}} (\bibinfo {year} {2013})}\BibitemShut {NoStop}%
\bibitem [{\citenamefont {Curtarolo}\ \emph {et~al.}(2012)\citenamefont {Curtarolo}, \citenamefont {Setyawan}, \citenamefont {Hart}, \citenamefont {Jahnatek}, \citenamefont {Chepulskii}, \citenamefont {Taylor}, \citenamefont {Wang}, \citenamefont {Xue}, \citenamefont {Yang}, \citenamefont {Levy} \emph {et~al.}}]{curtarolo2012aflow}%
  \BibitemOpen
  \bibfield  {author} {\bibinfo {author} {\bibfnamefont {S.}~\bibnamefont {Curtarolo}}, \bibinfo {author} {\bibfnamefont {W.}~\bibnamefont {Setyawan}}, \bibinfo {author} {\bibfnamefont {G.~L.}\ \bibnamefont {Hart}}, \bibinfo {author} {\bibfnamefont {M.}~\bibnamefont {Jahnatek}}, \bibinfo {author} {\bibfnamefont {R.~V.}\ \bibnamefont {Chepulskii}}, \bibinfo {author} {\bibfnamefont {R.~H.}\ \bibnamefont {Taylor}}, \bibinfo {author} {\bibfnamefont {S.}~\bibnamefont {Wang}}, \bibinfo {author} {\bibfnamefont {J.}~\bibnamefont {Xue}}, \bibinfo {author} {\bibfnamefont {K.}~\bibnamefont {Yang}}, \bibinfo {author} {\bibfnamefont {O.}~\bibnamefont {Levy}}, \emph {et~al.},\ }\bibfield  {title} {\bibinfo {title} {{AFLOW: An automatic framework for high-throughput materials discovery}},\ }\href@noop {} {\bibfield  {journal} {\bibinfo  {journal} {Comput. Mater. Sci.}\ }\textbf {\bibinfo {volume} {58}},\ \bibinfo {pages} {218} (\bibinfo {year} {2012})}\BibitemShut {NoStop}%
\bibitem [{\citenamefont {Montoya}\ \emph {et~al.}(2020)\citenamefont {Montoya}, \citenamefont {Winther}, \citenamefont {Flores}, \citenamefont {Bligaard}, \citenamefont {Hummelsh{\o}j},\ and\ \citenamefont {Aykol}}]{montoya2020autonomous}%
  \BibitemOpen
  \bibfield  {author} {\bibinfo {author} {\bibfnamefont {J.~H.}\ \bibnamefont {Montoya}}, \bibinfo {author} {\bibfnamefont {K.~T.}\ \bibnamefont {Winther}}, \bibinfo {author} {\bibfnamefont {R.~A.}\ \bibnamefont {Flores}}, \bibinfo {author} {\bibfnamefont {T.}~\bibnamefont {Bligaard}}, \bibinfo {author} {\bibfnamefont {J.~S.}\ \bibnamefont {Hummelsh{\o}j}},\ and\ \bibinfo {author} {\bibfnamefont {M.}~\bibnamefont {Aykol}},\ }\bibfield  {title} {\bibinfo {title} {{Autonomous intelligent agents for accelerated materials discovery}},\ }\href@noop {} {\bibfield  {journal} {\bibinfo  {journal} {Chem. Sci.}\ }\textbf {\bibinfo {volume} {11}},\ \bibinfo {pages} {8517} (\bibinfo {year} {2020})}\BibitemShut {NoStop}%
\bibitem [{\citenamefont {Ye}\ \emph {et~al.}(2022)\citenamefont {Ye}, \citenamefont {Lei}, \citenamefont {Aykol},\ and\ \citenamefont {Montoya}}]{ye2022novel}%
  \BibitemOpen
  \bibfield  {author} {\bibinfo {author} {\bibfnamefont {W.}~\bibnamefont {Ye}}, \bibinfo {author} {\bibfnamefont {X.}~\bibnamefont {Lei}}, \bibinfo {author} {\bibfnamefont {M.}~\bibnamefont {Aykol}},\ and\ \bibinfo {author} {\bibfnamefont {J.~H.}\ \bibnamefont {Montoya}},\ }\bibfield  {title} {\bibinfo {title} {{Novel inorganic crystal structures predicted using autonomous simulation agents}},\ }\href@noop {} {\bibfield  {journal} {\bibinfo  {journal} {Sci. Data}\ }\textbf {\bibinfo {volume} {9}},\ \bibinfo {pages} {302} (\bibinfo {year} {2022})}\BibitemShut {NoStop}%
\bibitem [{\citenamefont {Merchant}\ \emph {et~al.}(2023)\citenamefont {Merchant}, \citenamefont {Batzner}, \citenamefont {Schoenholz}, \citenamefont {Aykol}, \citenamefont {Cheon},\ and\ \citenamefont {Cubuk}}]{merchant2023scaling}%
  \BibitemOpen
  \bibfield  {author} {\bibinfo {author} {\bibfnamefont {A.}~\bibnamefont {Merchant}}, \bibinfo {author} {\bibfnamefont {S.}~\bibnamefont {Batzner}}, \bibinfo {author} {\bibfnamefont {S.~S.}\ \bibnamefont {Schoenholz}}, \bibinfo {author} {\bibfnamefont {M.}~\bibnamefont {Aykol}}, \bibinfo {author} {\bibfnamefont {G.}~\bibnamefont {Cheon}},\ and\ \bibinfo {author} {\bibfnamefont {E.~D.}\ \bibnamefont {Cubuk}},\ }\bibfield  {title} {\bibinfo {title} {{Scaling deep learning for materials discovery}},\ }\href@noop {} {\bibfield  {journal} {\bibinfo  {journal} {Nature}\ }\textbf {\bibinfo {volume} {624}},\ \bibinfo {pages} {80} (\bibinfo {year} {2023})}\BibitemShut {NoStop}%
\bibitem [{\citenamefont {Aykol}\ \emph {et~al.}(2018)\citenamefont {Aykol}, \citenamefont {Dwaraknath}, \citenamefont {Sun},\ and\ \citenamefont {Persson}}]{aykol2018thermodynamic}%
  \BibitemOpen
  \bibfield  {author} {\bibinfo {author} {\bibfnamefont {M.}~\bibnamefont {Aykol}}, \bibinfo {author} {\bibfnamefont {S.~S.}\ \bibnamefont {Dwaraknath}}, \bibinfo {author} {\bibfnamefont {W.}~\bibnamefont {Sun}},\ and\ \bibinfo {author} {\bibfnamefont {K.~A.}\ \bibnamefont {Persson}},\ }\bibfield  {title} {\bibinfo {title} {{Thermodynamic limit for synthesis of metastable inorganic materials}},\ }\href@noop {} {\bibfield  {journal} {\bibinfo  {journal} {Sci. Adv.}\ }\textbf {\bibinfo {volume} {4}},\ \bibinfo {pages} {eaaq0148} (\bibinfo {year} {2018})}\BibitemShut {NoStop}%
\bibitem [{\citenamefont {Montoya}\ \emph {et~al.}(2024)\citenamefont {Montoya}, \citenamefont {Grimley}, \citenamefont {Aykol}, \citenamefont {Ophus}, \citenamefont {Sternlicht}, \citenamefont {Savitzky}, \citenamefont {Minor}, \citenamefont {Torrisi}, \citenamefont {Goedjen}, \citenamefont {Chung} \emph {et~al.}}]{montoya2024ai}%
  \BibitemOpen
  \bibfield  {author} {\bibinfo {author} {\bibfnamefont {J.~H.}\ \bibnamefont {Montoya}}, \bibinfo {author} {\bibfnamefont {C.}~\bibnamefont {Grimley}}, \bibinfo {author} {\bibfnamefont {M.}~\bibnamefont {Aykol}}, \bibinfo {author} {\bibfnamefont {C.}~\bibnamefont {Ophus}}, \bibinfo {author} {\bibfnamefont {H.}~\bibnamefont {Sternlicht}}, \bibinfo {author} {\bibfnamefont {B.~H.}\ \bibnamefont {Savitzky}}, \bibinfo {author} {\bibfnamefont {A.~M.}\ \bibnamefont {Minor}}, \bibinfo {author} {\bibfnamefont {S.~B.}\ \bibnamefont {Torrisi}}, \bibinfo {author} {\bibfnamefont {J.}~\bibnamefont {Goedjen}}, \bibinfo {author} {\bibfnamefont {C.-C.}\ \bibnamefont {Chung}}, \emph {et~al.},\ }\bibfield  {title} {\bibinfo {title} {{How the AI-assisted discovery and synthesis of a ternary oxide highlights capability gaps in materials science}},\ }\href@noop {} {\bibfield  {journal} {\bibinfo  {journal} {Chem. Sci.}\ }\textbf {\bibinfo {volume} {15}},\ \bibinfo {pages} {5660} (\bibinfo {year} {2024})}\BibitemShut {NoStop}%
\bibitem [{\citenamefont {Singh}\ \emph {et~al.}(2017)\citenamefont {Singh}, \citenamefont {Zhou}, \citenamefont {Shinde}, \citenamefont {Suram}, \citenamefont {Montoya}, \citenamefont {Winston}, \citenamefont {Gregoire},\ and\ \citenamefont {Persson}}]{singh2017electrochemical}%
  \BibitemOpen
  \bibfield  {author} {\bibinfo {author} {\bibfnamefont {A.~K.}\ \bibnamefont {Singh}}, \bibinfo {author} {\bibfnamefont {L.}~\bibnamefont {Zhou}}, \bibinfo {author} {\bibfnamefont {A.}~\bibnamefont {Shinde}}, \bibinfo {author} {\bibfnamefont {S.~K.}\ \bibnamefont {Suram}}, \bibinfo {author} {\bibfnamefont {J.~H.}\ \bibnamefont {Montoya}}, \bibinfo {author} {\bibfnamefont {D.}~\bibnamefont {Winston}}, \bibinfo {author} {\bibfnamefont {J.~M.}\ \bibnamefont {Gregoire}},\ and\ \bibinfo {author} {\bibfnamefont {K.~A.}\ \bibnamefont {Persson}},\ }\bibfield  {title} {\bibinfo {title} {{Electrochemical stability of metastable materials}},\ }\href@noop {} {\bibfield  {journal} {\bibinfo  {journal} {Chem. Mater.}\ }\textbf {\bibinfo {volume} {29}},\ \bibinfo {pages} {10159} (\bibinfo {year} {2017})}\BibitemShut {NoStop}%
\bibitem [{\citenamefont {Lun}\ \emph {et~al.}(2021)\citenamefont {Lun}, \citenamefont {Ouyang}, \citenamefont {Kwon}, \citenamefont {Ha}, \citenamefont {Foley}, \citenamefont {Huang}, \citenamefont {Cai}, \citenamefont {Kim}, \citenamefont {Balasubramanian}, \citenamefont {Sun} \emph {et~al.}}]{lun2021cation}%
  \BibitemOpen
  \bibfield  {author} {\bibinfo {author} {\bibfnamefont {Z.}~\bibnamefont {Lun}}, \bibinfo {author} {\bibfnamefont {B.}~\bibnamefont {Ouyang}}, \bibinfo {author} {\bibfnamefont {D.-H.}\ \bibnamefont {Kwon}}, \bibinfo {author} {\bibfnamefont {Y.}~\bibnamefont {Ha}}, \bibinfo {author} {\bibfnamefont {E.~E.}\ \bibnamefont {Foley}}, \bibinfo {author} {\bibfnamefont {T.-Y.}\ \bibnamefont {Huang}}, \bibinfo {author} {\bibfnamefont {Z.}~\bibnamefont {Cai}}, \bibinfo {author} {\bibfnamefont {H.}~\bibnamefont {Kim}}, \bibinfo {author} {\bibfnamefont {M.}~\bibnamefont {Balasubramanian}}, \bibinfo {author} {\bibfnamefont {Y.}~\bibnamefont {Sun}}, \emph {et~al.},\ }\bibfield  {title} {\bibinfo {title} {{Cation-disordered rocksalt-type high-entropy cathodes for Li-ion batteries}},\ }\href@noop {} {\bibfield  {journal} {\bibinfo  {journal} {Nat. Mater.}\ }\textbf {\bibinfo {volume} {20}},\ \bibinfo {pages} {214} (\bibinfo {year} {2021})}\BibitemShut {NoStop}%
\bibitem [{\citenamefont {Divilov}\ \emph {et~al.}(2024)\citenamefont {Divilov}, \citenamefont {Eckert}, \citenamefont {Hicks}, \citenamefont {Oses}, \citenamefont {Toher}, \citenamefont {Friedrich}, \citenamefont {Esters}, \citenamefont {Mehl}, \citenamefont {Zettel}, \citenamefont {Lederer} \emph {et~al.}}]{divilov2024disordered}%
  \BibitemOpen
  \bibfield  {author} {\bibinfo {author} {\bibfnamefont {S.}~\bibnamefont {Divilov}}, \bibinfo {author} {\bibfnamefont {H.}~\bibnamefont {Eckert}}, \bibinfo {author} {\bibfnamefont {D.}~\bibnamefont {Hicks}}, \bibinfo {author} {\bibfnamefont {C.}~\bibnamefont {Oses}}, \bibinfo {author} {\bibfnamefont {C.}~\bibnamefont {Toher}}, \bibinfo {author} {\bibfnamefont {R.}~\bibnamefont {Friedrich}}, \bibinfo {author} {\bibfnamefont {M.}~\bibnamefont {Esters}}, \bibinfo {author} {\bibfnamefont {M.~J.}\ \bibnamefont {Mehl}}, \bibinfo {author} {\bibfnamefont {A.~C.}\ \bibnamefont {Zettel}}, \bibinfo {author} {\bibfnamefont {Y.}~\bibnamefont {Lederer}}, \emph {et~al.},\ }\bibfield  {title} {\bibinfo {title} {{Disordered enthalpy--entropy descriptor for high-entropy ceramics discovery}},\ }\href@noop {} {\bibfield  {journal} {\bibinfo  {journal} {Nature}\ }\textbf {\bibinfo {volume} {625}},\ \bibinfo {pages} {66} (\bibinfo {year} {2024})}\BibitemShut {NoStop}%
\bibitem [{\citenamefont {Ong}\ \emph {et~al.}(2008)\citenamefont {Ong}, \citenamefont {Wang}, \citenamefont {Kang},\ and\ \citenamefont {Ceder}}]{ong2008li}%
  \BibitemOpen
  \bibfield  {author} {\bibinfo {author} {\bibfnamefont {S.~P.}\ \bibnamefont {Ong}}, \bibinfo {author} {\bibfnamefont {L.}~\bibnamefont {Wang}}, \bibinfo {author} {\bibfnamefont {B.}~\bibnamefont {Kang}},\ and\ \bibinfo {author} {\bibfnamefont {G.}~\bibnamefont {Ceder}},\ }\bibfield  {title} {\bibinfo {title} {{Li-Fe-P-O$_2$ phase diagram from first principles calculations}},\ }\href@noop {} {\bibfield  {journal} {\bibinfo  {journal} {Chem. Mater.}\ }\textbf {\bibinfo {volume} {20}},\ \bibinfo {pages} {1798} (\bibinfo {year} {2008})}\BibitemShut {NoStop}%
\bibitem [{\citenamefont {Kirklin}\ \emph {et~al.}(2013)\citenamefont {Kirklin}, \citenamefont {Meredig},\ and\ \citenamefont {Wolverton}}]{kirklin2013high}%
  \BibitemOpen
  \bibfield  {author} {\bibinfo {author} {\bibfnamefont {S.}~\bibnamefont {Kirklin}}, \bibinfo {author} {\bibfnamefont {B.}~\bibnamefont {Meredig}},\ and\ \bibinfo {author} {\bibfnamefont {C.}~\bibnamefont {Wolverton}},\ }\bibfield  {title} {\bibinfo {title} {{High-throughput computational screening of new Li-ion battery anode materials}},\ }\href@noop {} {\bibfield  {journal} {\bibinfo  {journal} {Adv. Energy Mater.}\ }\textbf {\bibinfo {volume} {3}},\ \bibinfo {pages} {252} (\bibinfo {year} {2013})}\BibitemShut {NoStop}%
\bibitem [{\citenamefont {Aykol}\ and\ \citenamefont {Wolverton}(2014)}]{aykol2014local}%
  \BibitemOpen
  \bibfield  {author} {\bibinfo {author} {\bibfnamefont {M.}~\bibnamefont {Aykol}}\ and\ \bibinfo {author} {\bibfnamefont {C.}~\bibnamefont {Wolverton}},\ }\bibfield  {title} {\bibinfo {title} {{Local environment dependent GGA+ U method for accurate thermochemistry of transition metal compounds}},\ }\href@noop {} {\bibfield  {journal} {\bibinfo  {journal} {Phys. Rev. B}\ }\textbf {\bibinfo {volume} {90}},\ \bibinfo {pages} {115105} (\bibinfo {year} {2014})}\BibitemShut {NoStop}%
\bibitem [{\citenamefont {Perdew}\ and\ \citenamefont {Zunger}(1981)}]{perdew1981self}%
  \BibitemOpen
  \bibfield  {author} {\bibinfo {author} {\bibfnamefont {J.~P.}\ \bibnamefont {Perdew}}\ and\ \bibinfo {author} {\bibfnamefont {A.}~\bibnamefont {Zunger}},\ }\bibfield  {title} {\bibinfo {title} {{Self-interaction correction to density-functional approximations for many-electron systems}},\ }\href@noop {} {\bibfield  {journal} {\bibinfo  {journal} {Phys. Rev. B}\ }\textbf {\bibinfo {volume} {23}},\ \bibinfo {pages} {5048} (\bibinfo {year} {1981})}\BibitemShut {NoStop}%
\bibitem [{\citenamefont {Zhou}\ \emph {et~al.}(2004)\citenamefont {Zhou}, \citenamefont {Cococcioni}, \citenamefont {Marianetti}, \citenamefont {Morgan},\ and\ \citenamefont {Ceder}}]{zhou2004first}%
  \BibitemOpen
  \bibfield  {author} {\bibinfo {author} {\bibfnamefont {F.}~\bibnamefont {Zhou}}, \bibinfo {author} {\bibfnamefont {M.}~\bibnamefont {Cococcioni}}, \bibinfo {author} {\bibfnamefont {C.~A.}\ \bibnamefont {Marianetti}}, \bibinfo {author} {\bibfnamefont {D.}~\bibnamefont {Morgan}},\ and\ \bibinfo {author} {\bibfnamefont {G.}~\bibnamefont {Ceder}},\ }\bibfield  {title} {\bibinfo {title} {{First-principles prediction of redox potentials in transition-metal compounds with LDA+ U}},\ }\href@noop {} {\bibfield  {journal} {\bibinfo  {journal} {Phys. Rev. B}\ }\textbf {\bibinfo {volume} {70}},\ \bibinfo {pages} {235121} (\bibinfo {year} {2004})}\BibitemShut {NoStop}%
\bibitem [{\citenamefont {Anisimov}\ \emph {et~al.}(1997)\citenamefont {Anisimov}, \citenamefont {Aryasetiawan},\ and\ \citenamefont {Lichtenstein}}]{anisimov1997first}%
  \BibitemOpen
  \bibfield  {author} {\bibinfo {author} {\bibfnamefont {V.~I.}\ \bibnamefont {Anisimov}}, \bibinfo {author} {\bibfnamefont {F.}~\bibnamefont {Aryasetiawan}},\ and\ \bibinfo {author} {\bibfnamefont {A.}~\bibnamefont {Lichtenstein}},\ }\bibfield  {title} {\bibinfo {title} {{First-principles calculations of the electronic structure and spectra of strongly correlated systems: the LDA+ U method}},\ }\href@noop {} {\bibfield  {journal} {\bibinfo  {journal} {J. Phys.: Condens. Matter}\ }\textbf {\bibinfo {volume} {9}},\ \bibinfo {pages} {767} (\bibinfo {year} {1997})}\BibitemShut {NoStop}%
\bibitem [{\citenamefont {Liechtenstein}\ \emph {et~al.}(1995)\citenamefont {Liechtenstein}, \citenamefont {Anisimov},\ and\ \citenamefont {Zaanen}}]{liechtenstein1995density}%
  \BibitemOpen
  \bibfield  {author} {\bibinfo {author} {\bibfnamefont {A.}~\bibnamefont {Liechtenstein}}, \bibinfo {author} {\bibfnamefont {V.~I.}\ \bibnamefont {Anisimov}},\ and\ \bibinfo {author} {\bibfnamefont {J.}~\bibnamefont {Zaanen}},\ }\bibfield  {title} {\bibinfo {title} {{Density-functional theory and strong interactions: Orbital ordering in Mott-Hubbard insulators}},\ }\href@noop {} {\bibfield  {journal} {\bibinfo  {journal} {Phys. Rev. B}\ }\textbf {\bibinfo {volume} {52}},\ \bibinfo {pages} {R5467} (\bibinfo {year} {1995})}\BibitemShut {NoStop}%
\bibitem [{\citenamefont {Himmetoglu}\ \emph {et~al.}(2014)\citenamefont {Himmetoglu}, \citenamefont {Floris}, \citenamefont {De~Gironcoli},\ and\ \citenamefont {Cococcioni}}]{himmetoglu2014hubbard}%
  \BibitemOpen
  \bibfield  {author} {\bibinfo {author} {\bibfnamefont {B.}~\bibnamefont {Himmetoglu}}, \bibinfo {author} {\bibfnamefont {A.}~\bibnamefont {Floris}}, \bibinfo {author} {\bibfnamefont {S.}~\bibnamefont {De~Gironcoli}},\ and\ \bibinfo {author} {\bibfnamefont {M.}~\bibnamefont {Cococcioni}},\ }\bibfield  {title} {\bibinfo {title} {{Hubbard-corrected DFT energy functionals: The LDA+ U description of correlated systems}},\ }\href@noop {} {\bibfield  {journal} {\bibinfo  {journal} {Int. J. Quantum Chem.}\ }\textbf {\bibinfo {volume} {114}},\ \bibinfo {pages} {14} (\bibinfo {year} {2014})}\BibitemShut {NoStop}%
\bibitem [{\citenamefont {Wang}\ \emph {et~al.}(2006)\citenamefont {Wang}, \citenamefont {Maxisch},\ and\ \citenamefont {Ceder}}]{wang2006oxidation}%
  \BibitemOpen
  \bibfield  {author} {\bibinfo {author} {\bibfnamefont {L.}~\bibnamefont {Wang}}, \bibinfo {author} {\bibfnamefont {T.}~\bibnamefont {Maxisch}},\ and\ \bibinfo {author} {\bibfnamefont {G.}~\bibnamefont {Ceder}},\ }\bibfield  {title} {\bibinfo {title} {{Oxidation energies of transition metal oxides within the GGA+ U framework}},\ }\href@noop {} {\bibfield  {journal} {\bibinfo  {journal} {Phys. Rev. B}\ }\textbf {\bibinfo {volume} {73}},\ \bibinfo {pages} {195107} (\bibinfo {year} {2006})}\BibitemShut {NoStop}%
\bibitem [{\citenamefont {Jain}\ \emph {et~al.}(2011)\citenamefont {Jain}, \citenamefont {Hautier}, \citenamefont {Ong}, \citenamefont {Moore}, \citenamefont {Fischer}, \citenamefont {Persson},\ and\ \citenamefont {Ceder}}]{jain2011formation}%
  \BibitemOpen
  \bibfield  {author} {\bibinfo {author} {\bibfnamefont {A.}~\bibnamefont {Jain}}, \bibinfo {author} {\bibfnamefont {G.}~\bibnamefont {Hautier}}, \bibinfo {author} {\bibfnamefont {S.~P.}\ \bibnamefont {Ong}}, \bibinfo {author} {\bibfnamefont {C.~J.}\ \bibnamefont {Moore}}, \bibinfo {author} {\bibfnamefont {C.~C.}\ \bibnamefont {Fischer}}, \bibinfo {author} {\bibfnamefont {K.~A.}\ \bibnamefont {Persson}},\ and\ \bibinfo {author} {\bibfnamefont {G.}~\bibnamefont {Ceder}},\ }\bibfield  {title} {\bibinfo {title} {{Formation enthalpies by mixing GGA and GGA+ U calculations}},\ }\href@noop {} {\bibfield  {journal} {\bibinfo  {journal} {Phys. Rev. B}\ }\textbf {\bibinfo {volume} {84}},\ \bibinfo {pages} {045115} (\bibinfo {year} {2011})}\BibitemShut {NoStop}%
\bibitem [{\citenamefont {Calderon}\ \emph {et~al.}(2015)\citenamefont {Calderon}, \citenamefont {Plata}, \citenamefont {Toher}, \citenamefont {Oses}, \citenamefont {Levy}, \citenamefont {Fornari}, \citenamefont {Natan}, \citenamefont {Mehl}, \citenamefont {Hart}, \citenamefont {Nardelli} \emph {et~al.}}]{calderon2015aflow}%
  \BibitemOpen
  \bibfield  {author} {\bibinfo {author} {\bibfnamefont {C.~E.}\ \bibnamefont {Calderon}}, \bibinfo {author} {\bibfnamefont {J.~J.}\ \bibnamefont {Plata}}, \bibinfo {author} {\bibfnamefont {C.}~\bibnamefont {Toher}}, \bibinfo {author} {\bibfnamefont {C.}~\bibnamefont {Oses}}, \bibinfo {author} {\bibfnamefont {O.}~\bibnamefont {Levy}}, \bibinfo {author} {\bibfnamefont {M.}~\bibnamefont {Fornari}}, \bibinfo {author} {\bibfnamefont {A.}~\bibnamefont {Natan}}, \bibinfo {author} {\bibfnamefont {M.~J.}\ \bibnamefont {Mehl}}, \bibinfo {author} {\bibfnamefont {G.}~\bibnamefont {Hart}}, \bibinfo {author} {\bibfnamefont {M.~B.}\ \bibnamefont {Nardelli}}, \emph {et~al.},\ }\bibfield  {title} {\bibinfo {title} {{The AFLOW standard for high-throughput materials science calculations}},\ }\href@noop {} {\bibfield  {journal} {\bibinfo  {journal} {Comput. Mater. Sci.}\ }\textbf {\bibinfo {volume} {108}},\ \bibinfo {pages} {233} (\bibinfo {year} {2015})}\BibitemShut {NoStop}%
\bibitem [{\citenamefont {Dudarev}\ \emph {et~al.}(1998)\citenamefont {Dudarev}, \citenamefont {Botton}, \citenamefont {Savrasov}, \citenamefont {Humphreys},\ and\ \citenamefont {Sutton}}]{dudarev1998electron}%
  \BibitemOpen
  \bibfield  {author} {\bibinfo {author} {\bibfnamefont {S.~L.}\ \bibnamefont {Dudarev}}, \bibinfo {author} {\bibfnamefont {G.~A.}\ \bibnamefont {Botton}}, \bibinfo {author} {\bibfnamefont {S.~Y.}\ \bibnamefont {Savrasov}}, \bibinfo {author} {\bibfnamefont {C.}~\bibnamefont {Humphreys}},\ and\ \bibinfo {author} {\bibfnamefont {A.~P.}\ \bibnamefont {Sutton}},\ }\bibfield  {title} {\bibinfo {title} {{Electron-energy-loss spectra and the structural stability of nickel oxide: An LSDA+ U study}},\ }\href@noop {} {\bibfield  {journal} {\bibinfo  {journal} {Phys. Rev. B}\ }\textbf {\bibinfo {volume} {57}},\ \bibinfo {pages} {1505} (\bibinfo {year} {1998})}\BibitemShut {NoStop}%
\bibitem [{\citenamefont {Vaugier}\ \emph {et~al.}(2012)\citenamefont {Vaugier}, \citenamefont {Jiang},\ and\ \citenamefont {Biermann}}]{vaugier2012hubbard}%
  \BibitemOpen
  \bibfield  {author} {\bibinfo {author} {\bibfnamefont {L.}~\bibnamefont {Vaugier}}, \bibinfo {author} {\bibfnamefont {H.}~\bibnamefont {Jiang}},\ and\ \bibinfo {author} {\bibfnamefont {S.}~\bibnamefont {Biermann}},\ }\bibfield  {title} {\bibinfo {title} {{Hubbard U and Hund exchange J in transition metal oxides: Screening versus localization trends from constrained random phase approximation}},\ }\href@noop {} {\bibfield  {journal} {\bibinfo  {journal} {Phys. Rev. B}\ }\textbf {\bibinfo {volume} {86}},\ \bibinfo {pages} {165105} (\bibinfo {year} {2012})}\BibitemShut {NoStop}%
\bibitem [{\citenamefont {Cococcioni}\ and\ \citenamefont {De~Gironcoli}(2005)}]{cococcioni2005linear}%
  \BibitemOpen
  \bibfield  {author} {\bibinfo {author} {\bibfnamefont {M.}~\bibnamefont {Cococcioni}}\ and\ \bibinfo {author} {\bibfnamefont {S.}~\bibnamefont {De~Gironcoli}},\ }\bibfield  {title} {\bibinfo {title} {{Linear response approach to the calculation of the effective interaction parameters in the LDA+ U method}},\ }\href@noop {} {\bibfield  {journal} {\bibinfo  {journal} {Phys. Rev. B}\ }\textbf {\bibinfo {volume} {71}},\ \bibinfo {pages} {035105} (\bibinfo {year} {2005})}\BibitemShut {NoStop}%
\bibitem [{\citenamefont {Lutfalla}\ \emph {et~al.}(2011)\citenamefont {Lutfalla}, \citenamefont {Shapovalov},\ and\ \citenamefont {Bell}}]{lutfalla2011calibration}%
  \BibitemOpen
  \bibfield  {author} {\bibinfo {author} {\bibfnamefont {S.}~\bibnamefont {Lutfalla}}, \bibinfo {author} {\bibfnamefont {V.}~\bibnamefont {Shapovalov}},\ and\ \bibinfo {author} {\bibfnamefont {A.~T.}\ \bibnamefont {Bell}},\ }\bibfield  {title} {\bibinfo {title} {{Calibration of the DFT/GGA+ U method for determination of reduction energies for transition and rare earth metal oxides of Ti, V, Mo, and Ce}},\ }\href@noop {} {\bibfield  {journal} {\bibinfo  {journal} {J. Chem. Theory Comput.}\ }\textbf {\bibinfo {volume} {7}},\ \bibinfo {pages} {2218} (\bibinfo {year} {2011})}\BibitemShut {NoStop}%
\bibitem [{\citenamefont {Capdevila-Cortada}\ \emph {et~al.}(2016)\citenamefont {Capdevila-Cortada}, \citenamefont {{\L}odziana},\ and\ \citenamefont {L{\'o}pez}}]{capdevila2016performance}%
  \BibitemOpen
  \bibfield  {author} {\bibinfo {author} {\bibfnamefont {M.}~\bibnamefont {Capdevila-Cortada}}, \bibinfo {author} {\bibfnamefont {Z.}~\bibnamefont {{\L}odziana}},\ and\ \bibinfo {author} {\bibfnamefont {N.}~\bibnamefont {L{\'o}pez}},\ }\bibfield  {title} {\bibinfo {title} {{Performance of DFT+ U approaches in the study of catalytic materials}},\ }\href@noop {} {\bibfield  {journal} {\bibinfo  {journal} {ACS Catal.}\ }\textbf {\bibinfo {volume} {6}},\ \bibinfo {pages} {8370} (\bibinfo {year} {2016})}\BibitemShut {NoStop}%
\bibitem [{\citenamefont {{Materials Project}}()}]{materials_project_hubbard_u}%
  \BibitemOpen
  \bibfield  {author} {\bibinfo {author} {\bibnamefont {{Materials Project}}},\ }\href@noop {} {\bibinfo {title} {{Hubbard U Values}}},\ \bibinfo {howpublished} {\url{https://docs.materialsproject.org/methodology/materials-methodology/calculation-details/gga+u-calculations/hubbard-u-values}}\BibitemShut {NoStop}%
\bibitem [{\citenamefont {Setyawan}\ \emph {et~al.}(2011)\citenamefont {Setyawan}, \citenamefont {Gaume}, \citenamefont {Lam}, \citenamefont {Feigelson},\ and\ \citenamefont {Curtarolo}}]{setyawan2011high}%
  \BibitemOpen
  \bibfield  {author} {\bibinfo {author} {\bibfnamefont {W.}~\bibnamefont {Setyawan}}, \bibinfo {author} {\bibfnamefont {R.~M.}\ \bibnamefont {Gaume}}, \bibinfo {author} {\bibfnamefont {S.}~\bibnamefont {Lam}}, \bibinfo {author} {\bibfnamefont {R.~S.}\ \bibnamefont {Feigelson}},\ and\ \bibinfo {author} {\bibfnamefont {S.}~\bibnamefont {Curtarolo}},\ }\bibfield  {title} {\bibinfo {title} {{High-throughput combinatorial database of electronic band structures for inorganic scintillator materials}},\ }\href@noop {} {\bibfield  {journal} {\bibinfo  {journal} {ACS Comb. Sci.}\ }\textbf {\bibinfo {volume} {13}},\ \bibinfo {pages} {382} (\bibinfo {year} {2011})}\BibitemShut {NoStop}%
\bibitem [{\citenamefont {Moore}\ \emph {et~al.}(2024)\citenamefont {Moore}, \citenamefont {Horton}, \citenamefont {Linscott}, \citenamefont {Ganose}, \citenamefont {Siron}, \citenamefont {O'Regan},\ and\ \citenamefont {Persson}}]{moore2024high}%
  \BibitemOpen
  \bibfield  {author} {\bibinfo {author} {\bibfnamefont {G.~C.}\ \bibnamefont {Moore}}, \bibinfo {author} {\bibfnamefont {M.~K.}\ \bibnamefont {Horton}}, \bibinfo {author} {\bibfnamefont {E.}~\bibnamefont {Linscott}}, \bibinfo {author} {\bibfnamefont {A.~M.}\ \bibnamefont {Ganose}}, \bibinfo {author} {\bibfnamefont {M.}~\bibnamefont {Siron}}, \bibinfo {author} {\bibfnamefont {D.~D.}\ \bibnamefont {O'Regan}},\ and\ \bibinfo {author} {\bibfnamefont {K.~A.}\ \bibnamefont {Persson}},\ }\bibfield  {title} {\bibinfo {title} {{High-throughput determination of Hubbard U and Hund J values for transition metal oxides via the linear response formalism}},\ }\href@noop {} {\bibfield  {journal} {\bibinfo  {journal} {Phys. Rev. Mater.}\ }\textbf {\bibinfo {volume} {8}},\ \bibinfo {pages} {014409} (\bibinfo {year} {2024})}\BibitemShut {NoStop}%
\bibitem [{\citenamefont {Grimme}\ \emph {et~al.}(2010)\citenamefont {Grimme}, \citenamefont {Antony}, \citenamefont {Ehrlich},\ and\ \citenamefont {Krieg}}]{grimme2010consistent}%
  \BibitemOpen
  \bibfield  {author} {\bibinfo {author} {\bibfnamefont {S.}~\bibnamefont {Grimme}}, \bibinfo {author} {\bibfnamefont {J.}~\bibnamefont {Antony}}, \bibinfo {author} {\bibfnamefont {S.}~\bibnamefont {Ehrlich}},\ and\ \bibinfo {author} {\bibfnamefont {H.}~\bibnamefont {Krieg}},\ }\bibfield  {title} {\bibinfo {title} {{A consistent and accurate ab initio parametrization of density functional dispersion correction (DFT-D) for the 94 elements H-Pu}},\ }\href@noop {} {\bibfield  {journal} {\bibinfo  {journal} {J. Chem. Phys.}\ }\textbf {\bibinfo {volume} {132}} (\bibinfo {year} {2010})}\BibitemShut {NoStop}%
\bibitem [{\citenamefont {Aykol}\ \emph {et~al.}(2015)\citenamefont {Aykol}, \citenamefont {Kim},\ and\ \citenamefont {Wolverton}}]{aykol2015van}%
  \BibitemOpen
  \bibfield  {author} {\bibinfo {author} {\bibfnamefont {M.}~\bibnamefont {Aykol}}, \bibinfo {author} {\bibfnamefont {S.}~\bibnamefont {Kim}},\ and\ \bibinfo {author} {\bibfnamefont {C.}~\bibnamefont {Wolverton}},\ }\bibfield  {title} {\bibinfo {title} {{Van der Waals interactions in layered lithium cobalt oxides}},\ }\href@noop {} {\bibfield  {journal} {\bibinfo  {journal} {J. Phys. Chem. C}\ }\textbf {\bibinfo {volume} {119}},\ \bibinfo {pages} {19053} (\bibinfo {year} {2015})}\BibitemShut {NoStop}%
\bibitem [{\citenamefont {Furness}\ \emph {et~al.}(2020)\citenamefont {Furness}, \citenamefont {Kaplan}, \citenamefont {Ning}, \citenamefont {Perdew},\ and\ \citenamefont {Sun}}]{furness2020accurate}%
  \BibitemOpen
  \bibfield  {author} {\bibinfo {author} {\bibfnamefont {J.~W.}\ \bibnamefont {Furness}}, \bibinfo {author} {\bibfnamefont {A.~D.}\ \bibnamefont {Kaplan}}, \bibinfo {author} {\bibfnamefont {J.}~\bibnamefont {Ning}}, \bibinfo {author} {\bibfnamefont {J.~P.}\ \bibnamefont {Perdew}},\ and\ \bibinfo {author} {\bibfnamefont {J.}~\bibnamefont {Sun}},\ }\bibfield  {title} {\bibinfo {title} {{Accurate and numerically efficient r2SCAN meta-generalized gradient approximation}},\ }\href@noop {} {\bibfield  {journal} {\bibinfo  {journal} {J. Phys. Chem. Lett.}\ }\textbf {\bibinfo {volume} {11}},\ \bibinfo {pages} {8208} (\bibinfo {year} {2020})}\BibitemShut {NoStop}%
\bibitem [{\citenamefont {Heyd}\ \emph {et~al.}(2003)\citenamefont {Heyd}, \citenamefont {Scuseria},\ and\ \citenamefont {Ernzerhof}}]{heyd2003hybrid}%
  \BibitemOpen
  \bibfield  {author} {\bibinfo {author} {\bibfnamefont {J.}~\bibnamefont {Heyd}}, \bibinfo {author} {\bibfnamefont {G.~E.}\ \bibnamefont {Scuseria}},\ and\ \bibinfo {author} {\bibfnamefont {M.}~\bibnamefont {Ernzerhof}},\ }\bibfield  {title} {\bibinfo {title} {{Hybrid functionals based on a screened Coulomb potential}},\ }\href@noop {} {\bibfield  {journal} {\bibinfo  {journal} {J. Chem. Phys.}\ }\textbf {\bibinfo {volume} {118}},\ \bibinfo {pages} {8207} (\bibinfo {year} {2003})}\BibitemShut {NoStop}%
\bibitem [{\citenamefont {Heyd}\ and\ \citenamefont {Scuseria}(2004)}]{heyd2004efficient}%
  \BibitemOpen
  \bibfield  {author} {\bibinfo {author} {\bibfnamefont {J.}~\bibnamefont {Heyd}}\ and\ \bibinfo {author} {\bibfnamefont {G.~E.}\ \bibnamefont {Scuseria}},\ }\bibfield  {title} {\bibinfo {title} {Efficient hybrid density functional calculations in solids: Assessment of the heyd--scuseria--ernzerhof screened coulomb hybrid functional},\ }\href@noop {} {\bibfield  {journal} {\bibinfo  {journal} {J. Chem. Phys.}\ }\textbf {\bibinfo {volume} {121}},\ \bibinfo {pages} {1187} (\bibinfo {year} {2004})}\BibitemShut {NoStop}%
\bibitem [{\citenamefont {Krukau}\ \emph {et~al.}(2006)\citenamefont {Krukau}, \citenamefont {Vydrov}, \citenamefont {Izmaylov},\ and\ \citenamefont {Scuseria}}]{krukau2006influence}%
  \BibitemOpen
  \bibfield  {author} {\bibinfo {author} {\bibfnamefont {A.~V.}\ \bibnamefont {Krukau}}, \bibinfo {author} {\bibfnamefont {O.~A.}\ \bibnamefont {Vydrov}}, \bibinfo {author} {\bibfnamefont {A.~F.}\ \bibnamefont {Izmaylov}},\ and\ \bibinfo {author} {\bibfnamefont {G.~E.}\ \bibnamefont {Scuseria}},\ }\bibfield  {title} {\bibinfo {title} {{Influence of the exchange screening parameter on the performance of screened hybrid functionals}},\ }\href@noop {} {\bibfield  {journal} {\bibinfo  {journal} {J. Chem. Phys.}\ }\textbf {\bibinfo {volume} {125}} (\bibinfo {year} {2006})}\BibitemShut {NoStop}%
\bibitem [{\citenamefont {Chevrier}\ \emph {et~al.}(2010)\citenamefont {Chevrier}, \citenamefont {Ong}, \citenamefont {Armiento}, \citenamefont {Chan},\ and\ \citenamefont {Ceder}}]{chevrier2010hybrid}%
  \BibitemOpen
  \bibfield  {author} {\bibinfo {author} {\bibfnamefont {V.~L.}\ \bibnamefont {Chevrier}}, \bibinfo {author} {\bibfnamefont {S.~P.}\ \bibnamefont {Ong}}, \bibinfo {author} {\bibfnamefont {R.}~\bibnamefont {Armiento}}, \bibinfo {author} {\bibfnamefont {M.~K.}\ \bibnamefont {Chan}},\ and\ \bibinfo {author} {\bibfnamefont {G.}~\bibnamefont {Ceder}},\ }\bibfield  {title} {\bibinfo {title} {{Hybrid density functional calculations of redox potentials and formation energies of transition metal compounds}},\ }\href@noop {} {\bibfield  {journal} {\bibinfo  {journal} {Phys. Rev. B}\ }\textbf {\bibinfo {volume} {82}},\ \bibinfo {pages} {075122} (\bibinfo {year} {2010})}\BibitemShut {NoStop}%
\bibitem [{\citenamefont {Seo}\ \emph {et~al.}(2015)\citenamefont {Seo}, \citenamefont {Urban},\ and\ \citenamefont {Ceder}}]{seo2015calibrating}%
  \BibitemOpen
  \bibfield  {author} {\bibinfo {author} {\bibfnamefont {D.-H.}\ \bibnamefont {Seo}}, \bibinfo {author} {\bibfnamefont {A.}~\bibnamefont {Urban}},\ and\ \bibinfo {author} {\bibfnamefont {G.}~\bibnamefont {Ceder}},\ }\bibfield  {title} {\bibinfo {title} {{Calibrating transition-metal energy levels and oxygen bands in first-principles calculations: Accurate prediction of redox potentials and charge transfer in lithium transition-metal oxides}},\ }\href@noop {} {\bibfield  {journal} {\bibinfo  {journal} {Phys. Rev. B}\ }\textbf {\bibinfo {volume} {92}},\ \bibinfo {pages} {115118} (\bibinfo {year} {2015})}\BibitemShut {NoStop}%
\bibitem [{\citenamefont {Kresse}\ and\ \citenamefont {Hafner}(1993)}]{kresse1993ab}%
  \BibitemOpen
  \bibfield  {author} {\bibinfo {author} {\bibfnamefont {G.}~\bibnamefont {Kresse}}\ and\ \bibinfo {author} {\bibfnamefont {J.}~\bibnamefont {Hafner}},\ }\bibfield  {title} {\bibinfo {title} {{Ab initio molecular dynamics for liquid metals}},\ }\href@noop {} {\bibfield  {journal} {\bibinfo  {journal} {Phys. Rev. B}\ }\textbf {\bibinfo {volume} {47}},\ \bibinfo {pages} {558} (\bibinfo {year} {1993})}\BibitemShut {NoStop}%
\bibitem [{\citenamefont {Kresse}\ and\ \citenamefont {Furthm{\"u}ller}(1996{\natexlab{a}})}]{kresse1996efficiency}%
  \BibitemOpen
  \bibfield  {author} {\bibinfo {author} {\bibfnamefont {G.}~\bibnamefont {Kresse}}\ and\ \bibinfo {author} {\bibfnamefont {J.}~\bibnamefont {Furthm{\"u}ller}},\ }\bibfield  {title} {\bibinfo {title} {{Efficiency of ab-initio total energy calculations for metals and semiconductors using a plane-wave basis set}},\ }\href@noop {} {\bibfield  {journal} {\bibinfo  {journal} {Comput. Mater. Sci.}\ }\textbf {\bibinfo {volume} {6}},\ \bibinfo {pages} {15} (\bibinfo {year} {1996}{\natexlab{a}})}\BibitemShut {NoStop}%
\bibitem [{\citenamefont {Kresse}\ and\ \citenamefont {Furthm{\"u}ller}(1996{\natexlab{b}})}]{kresse1996efficient}%
  \BibitemOpen
  \bibfield  {author} {\bibinfo {author} {\bibfnamefont {G.}~\bibnamefont {Kresse}}\ and\ \bibinfo {author} {\bibfnamefont {J.}~\bibnamefont {Furthm{\"u}ller}},\ }\bibfield  {title} {\bibinfo {title} {{Efficient iterative schemes for ab initio total-energy calculations using a plane-wave basis set}},\ }\href@noop {} {\bibfield  {journal} {\bibinfo  {journal} {Phys. Rev. B}\ }\textbf {\bibinfo {volume} {54}},\ \bibinfo {pages} {11169} (\bibinfo {year} {1996}{\natexlab{b}})}\BibitemShut {NoStop}%
\bibitem [{\citenamefont {Bl{\"o}chl}(1994)}]{blochl1994projector}%
  \BibitemOpen
  \bibfield  {author} {\bibinfo {author} {\bibfnamefont {P.~E.}\ \bibnamefont {Bl{\"o}chl}},\ }\bibfield  {title} {\bibinfo {title} {{Projector augmented-wave method}},\ }\href@noop {} {\bibfield  {journal} {\bibinfo  {journal} {Phys. Rev. B}\ }\textbf {\bibinfo {volume} {50}},\ \bibinfo {pages} {17953} (\bibinfo {year} {1994})}\BibitemShut {NoStop}%
\bibitem [{\citenamefont {Kresse}\ and\ \citenamefont {Joubert}(1999)}]{kresse1999ultrasoft}%
  \BibitemOpen
  \bibfield  {author} {\bibinfo {author} {\bibfnamefont {G.}~\bibnamefont {Kresse}}\ and\ \bibinfo {author} {\bibfnamefont {D.}~\bibnamefont {Joubert}},\ }\bibfield  {title} {\bibinfo {title} {{From ultrasoft pseudopotentials to the projector augmented-wave method}},\ }\href@noop {} {\bibfield  {journal} {\bibinfo  {journal} {Phys. Rev. B}\ }\textbf {\bibinfo {volume} {59}},\ \bibinfo {pages} {1758} (\bibinfo {year} {1999})}\BibitemShut {NoStop}%
\bibitem [{\citenamefont {Grimme}\ \emph {et~al.}(2011)\citenamefont {Grimme}, \citenamefont {Ehrlich},\ and\ \citenamefont {Goerigk}}]{grimme2011effect}%
  \BibitemOpen
  \bibfield  {author} {\bibinfo {author} {\bibfnamefont {S.}~\bibnamefont {Grimme}}, \bibinfo {author} {\bibfnamefont {S.}~\bibnamefont {Ehrlich}},\ and\ \bibinfo {author} {\bibfnamefont {L.}~\bibnamefont {Goerigk}},\ }\bibfield  {title} {\bibinfo {title} {{Effect of the damping function in dispersion corrected density functional theory}},\ }\href@noop {} {\bibfield  {journal} {\bibinfo  {journal} {J. Comput. Chem.}\ }\textbf {\bibinfo {volume} {32}},\ \bibinfo {pages} {1456} (\bibinfo {year} {2011})}\BibitemShut {NoStop}%
\bibitem [{\citenamefont {Bl{\"o}chl}\ \emph {et~al.}(1994)\citenamefont {Bl{\"o}chl}, \citenamefont {Jepsen},\ and\ \citenamefont {Andersen}}]{blochl1994improved}%
  \BibitemOpen
  \bibfield  {author} {\bibinfo {author} {\bibfnamefont {P.~E.}\ \bibnamefont {Bl{\"o}chl}}, \bibinfo {author} {\bibfnamefont {O.}~\bibnamefont {Jepsen}},\ and\ \bibinfo {author} {\bibfnamefont {O.~K.}\ \bibnamefont {Andersen}},\ }\bibfield  {title} {\bibinfo {title} {{Improved tetrahedron method for Brillouin-zone integrations}},\ }\href@noop {} {\bibfield  {journal} {\bibinfo  {journal} {Phys. Rev. B}\ }\textbf {\bibinfo {volume} {49}},\ \bibinfo {pages} {16223} (\bibinfo {year} {1994})}\BibitemShut {NoStop}%
\bibitem [{\citenamefont {Togo}\ \emph {et~al.}(2023)\citenamefont {Togo}, \citenamefont {Chaput}, \citenamefont {Tadano},\ and\ \citenamefont {Tanaka}}]{phonopy-phono3py-JPCM}%
  \BibitemOpen
  \bibfield  {author} {\bibinfo {author} {\bibfnamefont {A.}~\bibnamefont {Togo}}, \bibinfo {author} {\bibfnamefont {L.}~\bibnamefont {Chaput}}, \bibinfo {author} {\bibfnamefont {T.}~\bibnamefont {Tadano}},\ and\ \bibinfo {author} {\bibfnamefont {I.}~\bibnamefont {Tanaka}},\ }\bibfield  {title} {\bibinfo {title} {{Implementation strategies in phonopy and phono3py}},\ }\href {https://doi.org/10.1088/1361-648X/acd831} {\bibfield  {journal} {\bibinfo  {journal} {J. Phys. Condens. Matter}\ }\textbf {\bibinfo {volume} {35}},\ \bibinfo {pages} {353001} (\bibinfo {year} {2023})}\BibitemShut {NoStop}%
\bibitem [{\citenamefont {Togo}(2023)}]{phonopy-phono3py-JPSJ}%
  \BibitemOpen
  \bibfield  {author} {\bibinfo {author} {\bibfnamefont {A.}~\bibnamefont {Togo}},\ }\bibfield  {title} {\bibinfo {title} {{First-principles Phonon Calculations with Phonopy and Phono3py}},\ }\href {https://doi.org/10.7566/JPSJ.92.012001} {\bibfield  {journal} {\bibinfo  {journal} {J. Phys. Soc. Jpn.}\ }\textbf {\bibinfo {volume} {92}},\ \bibinfo {pages} {012001} (\bibinfo {year} {2023})}\BibitemShut {NoStop}%
\bibitem [{\citenamefont {Spooner}\ \emph {et~al.}(2024)\citenamefont {Spooner}, \citenamefont {Einhorn}, \citenamefont {Davies},\ and\ \citenamefont {Scanlon}}]{spooner2024thermoparser}%
  \BibitemOpen
  \bibfield  {author} {\bibinfo {author} {\bibfnamefont {K.~B.}\ \bibnamefont {Spooner}}, \bibinfo {author} {\bibfnamefont {M.}~\bibnamefont {Einhorn}}, \bibinfo {author} {\bibfnamefont {D.~W.}\ \bibnamefont {Davies}},\ and\ \bibinfo {author} {\bibfnamefont {D.~O.}\ \bibnamefont {Scanlon}},\ }\bibfield  {title} {\bibinfo {title} {{ThermoParser: Streamlined analysis of thermoelectric properties}},\ }\href@noop {} {\bibfield  {journal} {\bibinfo  {journal} {J. Open Source Softw.}\ }\textbf {\bibinfo {volume} {9}},\ \bibinfo {pages} {6340} (\bibinfo {year} {2024})}\BibitemShut {NoStop}%
\bibitem [{\citenamefont {Zunger}\ \emph {et~al.}(1990)\citenamefont {Zunger}, \citenamefont {Wei}, \citenamefont {Ferreira},\ and\ \citenamefont {Bernard}}]{zunger1990special}%
  \BibitemOpen
  \bibfield  {author} {\bibinfo {author} {\bibfnamefont {A.}~\bibnamefont {Zunger}}, \bibinfo {author} {\bibfnamefont {S.-H.}\ \bibnamefont {Wei}}, \bibinfo {author} {\bibfnamefont {L.}~\bibnamefont {Ferreira}},\ and\ \bibinfo {author} {\bibfnamefont {J.~E.}\ \bibnamefont {Bernard}},\ }\bibfield  {title} {\bibinfo {title} {{Special quasirandom structures}},\ }\href@noop {} {\bibfield  {journal} {\bibinfo  {journal} {Phys. Rev. Lett.}\ }\textbf {\bibinfo {volume} {65}},\ \bibinfo {pages} {353} (\bibinfo {year} {1990})}\BibitemShut {NoStop}%
\bibitem [{\citenamefont {{\AA}ngqvist}\ \emph {et~al.}(2019)\citenamefont {{\AA}ngqvist}, \citenamefont {Mu{\~n}oz}, \citenamefont {Rahm}, \citenamefont {Fransson}, \citenamefont {Durniak}, \citenamefont {Rozyczko}, \citenamefont {Rod},\ and\ \citenamefont {Erhart}}]{aangqvist2019icet}%
  \BibitemOpen
  \bibfield  {author} {\bibinfo {author} {\bibfnamefont {M.}~\bibnamefont {{\AA}ngqvist}}, \bibinfo {author} {\bibfnamefont {W.~A.}\ \bibnamefont {Mu{\~n}oz}}, \bibinfo {author} {\bibfnamefont {J.~M.}\ \bibnamefont {Rahm}}, \bibinfo {author} {\bibfnamefont {E.}~\bibnamefont {Fransson}}, \bibinfo {author} {\bibfnamefont {C.}~\bibnamefont {Durniak}}, \bibinfo {author} {\bibfnamefont {P.}~\bibnamefont {Rozyczko}}, \bibinfo {author} {\bibfnamefont {T.~H.}\ \bibnamefont {Rod}},\ and\ \bibinfo {author} {\bibfnamefont {P.}~\bibnamefont {Erhart}},\ }\bibfield  {title} {\bibinfo {title} {{ICET--a Python library for constructing and sampling alloy cluster expansions}},\ }\href@noop {} {\bibfield  {journal} {\bibinfo  {journal} {Adv. Theory Simul.}\ }\textbf {\bibinfo {volume} {2}},\ \bibinfo {pages} {1900015} (\bibinfo {year} {2019})}\BibitemShut {NoStop}%
\bibitem [{\citenamefont {Van De~Walle}(2009)}]{van2009multicomponent}%
  \BibitemOpen
  \bibfield  {author} {\bibinfo {author} {\bibfnamefont {A.}~\bibnamefont {Van De~Walle}},\ }\bibfield  {title} {\bibinfo {title} {{Multicomponent multisublattice alloys, nonconfigurational entropy and other additions to the Alloy Theoretic Automated Toolkit}},\ }\href@noop {} {\bibfield  {journal} {\bibinfo  {journal} {Calphad}\ }\textbf {\bibinfo {volume} {33}},\ \bibinfo {pages} {266} (\bibinfo {year} {2009})}\BibitemShut {NoStop}%
\bibitem [{\citenamefont {Van~de Walle}\ \emph {et~al.}(2013)\citenamefont {Van~de Walle}, \citenamefont {Tiwary}, \citenamefont {De~Jong}, \citenamefont {Olmsted}, \citenamefont {Asta}, \citenamefont {Dick}, \citenamefont {Shin}, \citenamefont {Wang}, \citenamefont {Chen},\ and\ \citenamefont {Liu}}]{van2013efficient}%
  \BibitemOpen
  \bibfield  {author} {\bibinfo {author} {\bibfnamefont {A.}~\bibnamefont {Van~de Walle}}, \bibinfo {author} {\bibfnamefont {P.}~\bibnamefont {Tiwary}}, \bibinfo {author} {\bibfnamefont {M.}~\bibnamefont {De~Jong}}, \bibinfo {author} {\bibfnamefont {D.}~\bibnamefont {Olmsted}}, \bibinfo {author} {\bibfnamefont {M.}~\bibnamefont {Asta}}, \bibinfo {author} {\bibfnamefont {A.}~\bibnamefont {Dick}}, \bibinfo {author} {\bibfnamefont {D.}~\bibnamefont {Shin}}, \bibinfo {author} {\bibfnamefont {Y.}~\bibnamefont {Wang}}, \bibinfo {author} {\bibfnamefont {L.-Q.}\ \bibnamefont {Chen}},\ and\ \bibinfo {author} {\bibfnamefont {Z.-K.}\ \bibnamefont {Liu}},\ }\bibfield  {title} {\bibinfo {title} {{Efficient stochastic generation of special quasirandom structures}},\ }\href@noop {} {\bibfield  {journal} {\bibinfo  {journal} {Calphad}\ }\textbf {\bibinfo {volume} {42}},\ \bibinfo {pages} {13} (\bibinfo {year} {2013})}\BibitemShut {NoStop}%
\bibitem [{\citenamefont {Vitoux}\ \emph {et~al.}(2020)\citenamefont {Vitoux}, \citenamefont {Guignard}, \citenamefont {Penin}, \citenamefont {Carlier}, \citenamefont {Darriet},\ and\ \citenamefont {Delmas}}]{vitoux2020namoo2}%
  \BibitemOpen
  \bibfield  {author} {\bibinfo {author} {\bibfnamefont {L.}~\bibnamefont {Vitoux}}, \bibinfo {author} {\bibfnamefont {M.}~\bibnamefont {Guignard}}, \bibinfo {author} {\bibfnamefont {N.}~\bibnamefont {Penin}}, \bibinfo {author} {\bibfnamefont {D.}~\bibnamefont {Carlier}}, \bibinfo {author} {\bibfnamefont {J.}~\bibnamefont {Darriet}},\ and\ \bibinfo {author} {\bibfnamefont {C.}~\bibnamefont {Delmas}},\ }\bibfield  {title} {\bibinfo {title} {{NaMoO$_2$: a layered oxide with molybdenum clusters}},\ }\href@noop {} {\bibfield  {journal} {\bibinfo  {journal} {Inorg. Chem.}\ }\textbf {\bibinfo {volume} {59}},\ \bibinfo {pages} {4015} (\bibinfo {year} {2020})}\BibitemShut {NoStop}%
\bibitem [{\citenamefont {Dungey}\ \emph {et~al.}(1998)\citenamefont {Dungey}, \citenamefont {Curtis},\ and\ \citenamefont {Penner-Hahn}}]{dungey1998structural}%
  \BibitemOpen
  \bibfield  {author} {\bibinfo {author} {\bibfnamefont {K.~E.}\ \bibnamefont {Dungey}}, \bibinfo {author} {\bibfnamefont {M.~D.}\ \bibnamefont {Curtis}},\ and\ \bibinfo {author} {\bibfnamefont {J.~E.}\ \bibnamefont {Penner-Hahn}},\ }\bibfield  {title} {\bibinfo {title} {{Structural characterization and thermal stability of MoS$_2$ intercalation compounds}},\ }\href@noop {} {\bibfield  {journal} {\bibinfo  {journal} {Chem. Mater.}\ }\textbf {\bibinfo {volume} {10}},\ \bibinfo {pages} {2152} (\bibinfo {year} {1998})}\BibitemShut {NoStop}%
\bibitem [{\citenamefont {Petkov}\ \emph {et~al.}(2002)\citenamefont {Petkov}, \citenamefont {Billinge}, \citenamefont {Larson}, \citenamefont {Mahanti}, \citenamefont {Vogt}, \citenamefont {Rangan},\ and\ \citenamefont {Kanatzidis}}]{petkov2002structure}%
  \BibitemOpen
  \bibfield  {author} {\bibinfo {author} {\bibfnamefont {V.}~\bibnamefont {Petkov}}, \bibinfo {author} {\bibfnamefont {S.}~\bibnamefont {Billinge}}, \bibinfo {author} {\bibfnamefont {P.}~\bibnamefont {Larson}}, \bibinfo {author} {\bibfnamefont {S.}~\bibnamefont {Mahanti}}, \bibinfo {author} {\bibfnamefont {T.}~\bibnamefont {Vogt}}, \bibinfo {author} {\bibfnamefont {K.}~\bibnamefont {Rangan}},\ and\ \bibinfo {author} {\bibfnamefont {M.~G.}\ \bibnamefont {Kanatzidis}},\ }\bibfield  {title} {\bibinfo {title} {{Structure of nanocrystalline materials using atomic pair distribution function analysis: Study of LiMoS$_2$}},\ }\href@noop {} {\bibfield  {journal} {\bibinfo  {journal} {Phys. Rev. B}\ }\textbf {\bibinfo {volume} {65}},\ \bibinfo {pages} {092105} (\bibinfo {year} {2002})}\BibitemShut {NoStop}%
\bibitem [{\citenamefont {Rocquefelte}\ \emph {et~al.}(2000)\citenamefont {Rocquefelte}, \citenamefont {Boucher}, \citenamefont {Gressier}, \citenamefont {Ouvrard}, \citenamefont {Blaha},\ and\ \citenamefont {Schwarz}}]{rocquefelte2000mo}%
  \BibitemOpen
  \bibfield  {author} {\bibinfo {author} {\bibfnamefont {X.}~\bibnamefont {Rocquefelte}}, \bibinfo {author} {\bibfnamefont {F.}~\bibnamefont {Boucher}}, \bibinfo {author} {\bibfnamefont {P.}~\bibnamefont {Gressier}}, \bibinfo {author} {\bibfnamefont {G.}~\bibnamefont {Ouvrard}}, \bibinfo {author} {\bibfnamefont {P.}~\bibnamefont {Blaha}},\ and\ \bibinfo {author} {\bibfnamefont {K.}~\bibnamefont {Schwarz}},\ }\bibfield  {title} {\bibinfo {title} {{Mo cluster formation in the intercalation compound LiMoS$_2$}},\ }\href@noop {} {\bibfield  {journal} {\bibinfo  {journal} {Phys. Rev. B}\ }\textbf {\bibinfo {volume} {62}},\ \bibinfo {pages} {2397} (\bibinfo {year} {2000})}\BibitemShut {NoStop}%
\bibitem [{\citenamefont {Wildervanck}\ and\ \citenamefont {Jellinek}(1971)}]{wildervanck1971dichalcogenides}%
  \BibitemOpen
  \bibfield  {author} {\bibinfo {author} {\bibfnamefont {J.}~\bibnamefont {Wildervanck}}\ and\ \bibinfo {author} {\bibfnamefont {F.}~\bibnamefont {Jellinek}},\ }\bibfield  {title} {\bibinfo {title} {{The dichalcogenides of technetium and rhenium}},\ }\href@noop {} {\bibfield  {journal} {\bibinfo  {journal} {J. Less-Common Met.}\ }\textbf {\bibinfo {volume} {24}},\ \bibinfo {pages} {73} (\bibinfo {year} {1971})}\BibitemShut {NoStop}%
\bibitem [{\citenamefont {Lamfers}\ \emph {et~al.}(1996)\citenamefont {Lamfers}, \citenamefont {Meetsma}, \citenamefont {Wiegers},\ and\ \citenamefont {De~Boer}}]{lamfers1996crystal}%
  \BibitemOpen
  \bibfield  {author} {\bibinfo {author} {\bibfnamefont {H.-J.}\ \bibnamefont {Lamfers}}, \bibinfo {author} {\bibfnamefont {A.}~\bibnamefont {Meetsma}}, \bibinfo {author} {\bibfnamefont {G.}~\bibnamefont {Wiegers}},\ and\ \bibinfo {author} {\bibfnamefont {J.}~\bibnamefont {De~Boer}},\ }\bibfield  {title} {\bibinfo {title} {{The crystal structure of some rhenium and technetium dichalcogenides}},\ }\href@noop {} {\bibfield  {journal} {\bibinfo  {journal} {J. Alloys Compd.}\ }\textbf {\bibinfo {volume} {241}},\ \bibinfo {pages} {34} (\bibinfo {year} {1996})}\BibitemShut {NoStop}%
\bibitem [{\citenamefont {Fang}\ \emph {et~al.}(1997)\citenamefont {Fang}, \citenamefont {Wiegers}, \citenamefont {Haas},\ and\ \citenamefont {De~Groot}}]{fang1997electronic}%
  \BibitemOpen
  \bibfield  {author} {\bibinfo {author} {\bibfnamefont {C.}~\bibnamefont {Fang}}, \bibinfo {author} {\bibfnamefont {G.}~\bibnamefont {Wiegers}}, \bibinfo {author} {\bibfnamefont {C.}~\bibnamefont {Haas}},\ and\ \bibinfo {author} {\bibfnamefont {R.}~\bibnamefont {De~Groot}},\ }\bibfield  {title} {\bibinfo {title} {{Electronic structures of ReS$_{2}$, ReSe$_{2}$, and TcS$_{2}$ in the real and the hypothetical undistorted structures}},\ }\href@noop {} {\bibfield  {journal} {\bibinfo  {journal} {J. Phys.: Condens. Matter}\ }\textbf {\bibinfo {volume} {9}},\ \bibinfo {pages} {4411} (\bibinfo {year} {1997})}\BibitemShut {NoStop}%
\bibitem [{\citenamefont {Peierls}(1996)}]{peierls1996quantum}%
  \BibitemOpen
  \bibfield  {author} {\bibinfo {author} {\bibfnamefont {R.~E.}\ \bibnamefont {Peierls}},\ }\href@noop {} {\emph {\bibinfo {title} {{Quantum theory of solids}}}}\ (\bibinfo  {publisher} {Clarendon Press},\ \bibinfo {year} {1996})\BibitemShut {NoStop}%
\bibitem [{\citenamefont {Canadell}\ \emph {et~al.}(1989)\citenamefont {Canadell}, \citenamefont {LeBeuze}, \citenamefont {El~Khalifa}, \citenamefont {Chevrel},\ and\ \citenamefont {Whangbo}}]{canadell1989origin}%
  \BibitemOpen
  \bibfield  {author} {\bibinfo {author} {\bibfnamefont {E.}~\bibnamefont {Canadell}}, \bibinfo {author} {\bibfnamefont {A.}~\bibnamefont {LeBeuze}}, \bibinfo {author} {\bibfnamefont {M.~A.}\ \bibnamefont {El~Khalifa}}, \bibinfo {author} {\bibfnamefont {R.}~\bibnamefont {Chevrel}},\ and\ \bibinfo {author} {\bibfnamefont {M.~H.}\ \bibnamefont {Whangbo}},\ }\bibfield  {title} {\bibinfo {title} {{Origin of metal clustering in transition-metal chalcogenide layers MX$_2$ (M= Nb, Ta, Mo, Re; X= S, Se)}},\ }\href@noop {} {\bibfield  {journal} {\bibinfo  {journal} {J. Am. Chem. Soc.}\ }\textbf {\bibinfo {volume} {111}},\ \bibinfo {pages} {3778} (\bibinfo {year} {1989})}\BibitemShut {NoStop}%
\bibitem [{\citenamefont {Hibble}\ \emph {et~al.}(1997)\citenamefont {Hibble}, \citenamefont {Fawcett},\ and\ \citenamefont {Hannon}}]{hibble1997true}%
  \BibitemOpen
  \bibfield  {author} {\bibinfo {author} {\bibfnamefont {S.~J.}\ \bibnamefont {Hibble}}, \bibinfo {author} {\bibfnamefont {I.~D.}\ \bibnamefont {Fawcett}},\ and\ \bibinfo {author} {\bibfnamefont {A.~C.}\ \bibnamefont {Hannon}},\ }\bibfield  {title} {\bibinfo {title} {{The true structure and metal- metal-bonded framework of LiMo$^{III}$O$_2$ determined from total neutron scattering}},\ }\href@noop {} {\bibfield  {journal} {\bibinfo  {journal} {Inorg. Chem.}\ }\textbf {\bibinfo {volume} {36}},\ \bibinfo {pages} {1749} (\bibinfo {year} {1997})}\BibitemShut {NoStop}%
\bibitem [{Note1()}]{Note1}%
  \BibitemOpen
  \bibinfo {note} {There is a possibility these instabilities arise from small numerical errors in the forces or supercell size effects. Some of the observed imaginary phonon modes occur at wavevectors that are incommensurate with the supercell and are thus not sampled exactly. We cannot rule out the possibility that these structures are instead dynamically stable but close to instability. Further confirmation of dynamic (in)stability would require relaxation of long period supercells that are prohibitively expensive to calculate. Nevertheless, small instabilities are unlikely to lead to large changes in energy after relaxation and thus do not significantly change our conclusions}\BibitemShut {NoStop}%
\bibitem [{\citenamefont {Tadano}\ and\ \citenamefont {Tsuneyuki}(2015)}]{PhysRevB.92.054301}%
  \BibitemOpen
  \bibfield  {author} {\bibinfo {author} {\bibfnamefont {T.}~\bibnamefont {Tadano}}\ and\ \bibinfo {author} {\bibfnamefont {S.}~\bibnamefont {Tsuneyuki}},\ }\bibfield  {title} {\bibinfo {title} {{Self-consistent phonon calculations of lattice dynamical properties in cubic ${\mathrm{SrTiO}}_{3}$ with first-principles anharmonic force constants}},\ }\href {https://doi.org/10.1103/PhysRevB.92.054301} {\bibfield  {journal} {\bibinfo  {journal} {Phys. Rev. B}\ }\textbf {\bibinfo {volume} {92}},\ \bibinfo {pages} {054301} (\bibinfo {year} {2015})}\BibitemShut {NoStop}%
\bibitem [{\citenamefont {Skelton}\ \emph {et~al.}(2016)\citenamefont {Skelton}, \citenamefont {Burton}, \citenamefont {Parker}, \citenamefont {Walsh}, \citenamefont {Kim}, \citenamefont {Soon}, \citenamefont {Buckeridge}, \citenamefont {Sokol}, \citenamefont {Catlow}, \citenamefont {Togo},\ and\ \citenamefont {Tanaka}}]{PhysRevLett.117.075502}%
  \BibitemOpen
  \bibfield  {author} {\bibinfo {author} {\bibfnamefont {J.~M.}\ \bibnamefont {Skelton}}, \bibinfo {author} {\bibfnamefont {L.~A.}\ \bibnamefont {Burton}}, \bibinfo {author} {\bibfnamefont {S.~C.}\ \bibnamefont {Parker}}, \bibinfo {author} {\bibfnamefont {A.}~\bibnamefont {Walsh}}, \bibinfo {author} {\bibfnamefont {C.-E.}\ \bibnamefont {Kim}}, \bibinfo {author} {\bibfnamefont {A.}~\bibnamefont {Soon}}, \bibinfo {author} {\bibfnamefont {J.}~\bibnamefont {Buckeridge}}, \bibinfo {author} {\bibfnamefont {A.~A.}\ \bibnamefont {Sokol}}, \bibinfo {author} {\bibfnamefont {C.~R.~A.}\ \bibnamefont {Catlow}}, \bibinfo {author} {\bibfnamefont {A.}~\bibnamefont {Togo}},\ and\ \bibinfo {author} {\bibfnamefont {I.}~\bibnamefont {Tanaka}},\ }\bibfield  {title} {\bibinfo {title} {{Anharmonicity in the High-Temperature $Cmcm$ Phase of SnSe: Soft Modes and Three-Phonon Interactions}},\ }\href {https://doi.org/10.1103/PhysRevLett.117.075502} {\bibfield  {journal} {\bibinfo  {journal} {Phys. Rev. Lett.}\ }\textbf {\bibinfo
  {volume} {117}},\ \bibinfo {pages} {075502} (\bibinfo {year} {2016})}\BibitemShut {NoStop}%
\bibitem [{\citenamefont {Xia}\ \emph {et~al.}(2020)\citenamefont {Xia}, \citenamefont {Ozoli\ifmmode \mbox{\c{n}}\else \c{n}\fi{}\ifmmode~\check{s}\else \v{s}\fi{}},\ and\ \citenamefont {Wolverton}}]{PhysRevLett.125.085901}%
  \BibitemOpen
  \bibfield  {author} {\bibinfo {author} {\bibfnamefont {Y.}~\bibnamefont {Xia}}, \bibinfo {author} {\bibfnamefont {V.}~\bibnamefont {Ozoli\ifmmode \mbox{\c{n}}\else \c{n}\fi{}\ifmmode~\check{s}\else \v{s}\fi{}}},\ and\ \bibinfo {author} {\bibfnamefont {C.}~\bibnamefont {Wolverton}},\ }\bibfield  {title} {\bibinfo {title} {{Microscopic Mechanisms of Glasslike Lattice Thermal Transport in Cubic ${\mathrm{Cu}}_{12}{\mathrm{Sb}}_{4}{\mathrm{S}}_{13}$ Tetrahedrites}},\ }\href {https://doi.org/10.1103/PhysRevLett.125.085901} {\bibfield  {journal} {\bibinfo  {journal} {Phys. Rev. Lett.}\ }\textbf {\bibinfo {volume} {125}},\ \bibinfo {pages} {085901} (\bibinfo {year} {2020})}\BibitemShut {NoStop}%
\bibitem [{\citenamefont {Aleandri}\ and\ \citenamefont {McCarley}(1988)}]{aleandri1988hexagonal}%
  \BibitemOpen
  \bibfield  {author} {\bibinfo {author} {\bibfnamefont {L.~E.}\ \bibnamefont {Aleandri}}\ and\ \bibinfo {author} {\bibfnamefont {R.~E.}\ \bibnamefont {McCarley}},\ }\bibfield  {title} {\bibinfo {title} {{Hexagonal lithium molybdate, LiMoO$_2$: a close-packed layered structure with infinite molybdenum-molybdenum-bonded sheets}},\ }\href@noop {} {\bibfield  {journal} {\bibinfo  {journal} {Inorg. Chem.}\ }\textbf {\bibinfo {volume} {27}},\ \bibinfo {pages} {1041} (\bibinfo {year} {1988})}\BibitemShut {NoStop}%
\bibitem [{\citenamefont {Barker}\ \emph {et~al.}(2003{\natexlab{a}})\citenamefont {Barker}, \citenamefont {Saidi},\ and\ \citenamefont {Swoyer}}]{barker2003lithium}%
  \BibitemOpen
  \bibfield  {author} {\bibinfo {author} {\bibfnamefont {J.}~\bibnamefont {Barker}}, \bibinfo {author} {\bibfnamefont {M.}~\bibnamefont {Saidi}},\ and\ \bibinfo {author} {\bibfnamefont {J.}~\bibnamefont {Swoyer}},\ }\bibfield  {title} {\bibinfo {title} {{Lithium insertion properties of the layered LiMoO$_2$ (R3m) made by a novel carbothermal reduction method}},\ }\href@noop {} {\bibfield  {journal} {\bibinfo  {journal} {Solid State Ion.}\ }\textbf {\bibinfo {volume} {158}},\ \bibinfo {pages} {261} (\bibinfo {year} {2003}{\natexlab{a}})}\BibitemShut {NoStop}%
\bibitem [{\citenamefont {Barker}\ \emph {et~al.}(2003{\natexlab{b}})\citenamefont {Barker}, \citenamefont {Saidi},\ and\ \citenamefont {Swoyer}}]{barker2003synthesis}%
  \BibitemOpen
  \bibfield  {author} {\bibinfo {author} {\bibfnamefont {J.}~\bibnamefont {Barker}}, \bibinfo {author} {\bibfnamefont {M.}~\bibnamefont {Saidi}},\ and\ \bibinfo {author} {\bibfnamefont {J.}~\bibnamefont {Swoyer}},\ }\bibfield  {title} {\bibinfo {title} {{Synthesis and Electrochemical Insertion Properties of the Layered Li$_x$MoO$_2$ Phases (x= 0.74, 0.85, and 1.00)}},\ }\href@noop {} {\bibfield  {journal} {\bibinfo  {journal} {Electrochem. Solid-State Lett.}\ }\textbf {\bibinfo {volume} {6}},\ \bibinfo {pages} {A252} (\bibinfo {year} {2003}{\natexlab{b}})}\BibitemShut {NoStop}%
\bibitem [{\citenamefont {Hibble}\ and\ \citenamefont {Fawcett}(1995)}]{hibble1995local}%
  \BibitemOpen
  \bibfield  {author} {\bibinfo {author} {\bibfnamefont {S.}~\bibnamefont {Hibble}}\ and\ \bibinfo {author} {\bibfnamefont {I.}~\bibnamefont {Fawcett}},\ }\bibfield  {title} {\bibinfo {title} {{Local order and metal-metal bonding in Li$_2$MoO$_3$, Li$_4$Mo$_3$O$_8$, LiMoO$_2$, and H$_2$MoO$_3$, determined from EXAFS studies}},\ }\href@noop {} {\bibfield  {journal} {\bibinfo  {journal} {Inorg. Chem.}\ }\textbf {\bibinfo {volume} {34}},\ \bibinfo {pages} {500} (\bibinfo {year} {1995})}\BibitemShut {NoStop}%
\bibitem [{\citenamefont {Ben-Kamel}\ \emph {et~al.}(2012)\citenamefont {Ben-Kamel}, \citenamefont {Amdouni}, \citenamefont {Groult}, \citenamefont {Mauger}, \citenamefont {Zaghib},\ and\ \citenamefont {Julien}}]{ben2012structural}%
  \BibitemOpen
  \bibfield  {author} {\bibinfo {author} {\bibfnamefont {K.}~\bibnamefont {Ben-Kamel}}, \bibinfo {author} {\bibfnamefont {N.}~\bibnamefont {Amdouni}}, \bibinfo {author} {\bibfnamefont {H.}~\bibnamefont {Groult}}, \bibinfo {author} {\bibfnamefont {A.}~\bibnamefont {Mauger}}, \bibinfo {author} {\bibfnamefont {K.}~\bibnamefont {Zaghib}},\ and\ \bibinfo {author} {\bibfnamefont {C.}~\bibnamefont {Julien}},\ }\bibfield  {title} {\bibinfo {title} {{Structural and electrochemical properties of LiMoO$_2$}},\ }\href@noop {} {\bibfield  {journal} {\bibinfo  {journal} {J. Power Sources}\ }\textbf {\bibinfo {volume} {202}},\ \bibinfo {pages} {314} (\bibinfo {year} {2012})}\BibitemShut {NoStop}%
\bibitem [{\citenamefont {Ramana}\ \emph {et~al.}(2021)\citenamefont {Ramana}, \citenamefont {Mauger},\ and\ \citenamefont {Julien}}]{ramana2021growth}%
  \BibitemOpen
  \bibfield  {author} {\bibinfo {author} {\bibfnamefont {C.}~\bibnamefont {Ramana}}, \bibinfo {author} {\bibfnamefont {A.}~\bibnamefont {Mauger}},\ and\ \bibinfo {author} {\bibfnamefont {C.}~\bibnamefont {Julien}},\ }\bibfield  {title} {\bibinfo {title} {{Growth, characterization and performance of bulk and nanoengineered molybdenum oxides for electrochemical energy storage and conversion}},\ }\href@noop {} {\bibfield  {journal} {\bibinfo  {journal} {Prog. Cryst. Growth Charact. Mater}\ }\textbf {\bibinfo {volume} {67}},\ \bibinfo {pages} {100533} (\bibinfo {year} {2021})}\BibitemShut {NoStop}%
\bibitem [{\citenamefont {Mikhailova}\ \emph {et~al.}(2011)\citenamefont {Mikhailova}, \citenamefont {Bramnik}, \citenamefont {Bramnik}, \citenamefont {Reichel}, \citenamefont {Oswald}, \citenamefont {Senyshyn}, \citenamefont {Trots},\ and\ \citenamefont {Ehrenberg}}]{mikhailova2011layered}%
  \BibitemOpen
  \bibfield  {author} {\bibinfo {author} {\bibfnamefont {D.}~\bibnamefont {Mikhailova}}, \bibinfo {author} {\bibfnamefont {N.}~\bibnamefont {Bramnik}}, \bibinfo {author} {\bibfnamefont {K.}~\bibnamefont {Bramnik}}, \bibinfo {author} {\bibfnamefont {P.}~\bibnamefont {Reichel}}, \bibinfo {author} {\bibfnamefont {S.}~\bibnamefont {Oswald}}, \bibinfo {author} {\bibfnamefont {A.}~\bibnamefont {Senyshyn}}, \bibinfo {author} {\bibfnamefont {D.}~\bibnamefont {Trots}},\ and\ \bibinfo {author} {\bibfnamefont {H.}~\bibnamefont {Ehrenberg}},\ }\bibfield  {title} {\bibinfo {title} {{Layered Li$_x$MoO$_2$ Phases with Different Composition for Electrochemical Application: Structural Considerations}},\ }\href@noop {} {\bibfield  {journal} {\bibinfo  {journal} {Chem. Mater.}\ }\textbf {\bibinfo {volume} {23}},\ \bibinfo {pages} {3429} (\bibinfo {year} {2011})}\BibitemShut {NoStop}%
\bibitem [{\citenamefont {Isaacs}\ \emph {et~al.}(2020)\citenamefont {Isaacs}, \citenamefont {Patel},\ and\ \citenamefont {Wolverton}}]{isaacs2020prediction}%
  \BibitemOpen
  \bibfield  {author} {\bibinfo {author} {\bibfnamefont {E.~B.}\ \bibnamefont {Isaacs}}, \bibinfo {author} {\bibfnamefont {S.}~\bibnamefont {Patel}},\ and\ \bibinfo {author} {\bibfnamefont {C.}~\bibnamefont {Wolverton}},\ }\bibfield  {title} {\bibinfo {title} {{Prediction of Li intercalation voltages in rechargeable battery cathode materials: Effects of exchange-correlation functional, van der Waals interactions, and Hubbard U}},\ }\href@noop {} {\bibfield  {journal} {\bibinfo  {journal} {Phys. Rev. Mater.}\ }\textbf {\bibinfo {volume} {4}},\ \bibinfo {pages} {065405} (\bibinfo {year} {2020})}\BibitemShut {NoStop}%
\bibitem [{\citenamefont {Perdew}\ \emph {et~al.}(1996)\citenamefont {Perdew}, \citenamefont {Ernzerhof},\ and\ \citenamefont {Burke}}]{perdew1996rationale}%
  \BibitemOpen
  \bibfield  {author} {\bibinfo {author} {\bibfnamefont {J.~P.}\ \bibnamefont {Perdew}}, \bibinfo {author} {\bibfnamefont {M.}~\bibnamefont {Ernzerhof}},\ and\ \bibinfo {author} {\bibfnamefont {K.}~\bibnamefont {Burke}},\ }\bibfield  {title} {\bibinfo {title} {{Rationale for mixing exact exchange with density functional approximations}},\ }\href@noop {} {\bibfield  {journal} {\bibinfo  {journal} {J. Chem. Phys.}\ }\textbf {\bibinfo {volume} {105}},\ \bibinfo {pages} {9982} (\bibinfo {year} {1996})}\BibitemShut {NoStop}%
\bibitem [{\citenamefont {Ringenbach}\ and\ \citenamefont {Kessler}(1969)}]{ringenbach1969hatterer}%
  \BibitemOpen
  \bibfield  {author} {\bibinfo {author} {\bibfnamefont {C.}~\bibnamefont {Ringenbach}}\ and\ \bibinfo {author} {\bibfnamefont {H.}~\bibnamefont {Kessler}},\ }\bibfield  {title} {\bibinfo {title} {{Hatterer. A. Un Nouveau Compos{\'e} Oxyg{\'e}n{\'e} du Molybd{\`e}ne NaMoO2. Propri{\'e}t{\'e}s Cristallographiques et Magn{\'e}tiques}},\ }\href@noop {} {\bibfield  {journal} {\bibinfo  {journal} {CR Acad. Sc. Paris, S{\'e}rie C}\ }\textbf {\bibinfo {volume} {269}},\ \bibinfo {pages} {1394} (\bibinfo {year} {1969})}\BibitemShut {NoStop}%
\bibitem [{\citenamefont {Brandt}\ \emph {et~al.}(1967)\citenamefont {Brandt}, \citenamefont {Skapski}, \citenamefont {Thom}, \citenamefont {Stoll}, \citenamefont {Eriksson},\ and\ \citenamefont {Blinc}}]{brandt1967refinement}%
  \BibitemOpen
  \bibfield  {author} {\bibinfo {author} {\bibfnamefont {B.~G.}\ \bibnamefont {Brandt}}, \bibinfo {author} {\bibfnamefont {A.}~\bibnamefont {Skapski}}, \bibinfo {author} {\bibfnamefont {E.}~\bibnamefont {Thom}}, \bibinfo {author} {\bibfnamefont {E.}~\bibnamefont {Stoll}}, \bibinfo {author} {\bibfnamefont {G.}~\bibnamefont {Eriksson}},\ and\ \bibinfo {author} {\bibfnamefont {R.}~\bibnamefont {Blinc}},\ }\bibfield  {title} {\bibinfo {title} {{A refinement of the crystal structure of molybdenum dioxide}},\ }\href@noop {} {\bibfield  {journal} {\bibinfo  {journal} {Acta Chem. Scand}\ }\textbf {\bibinfo {volume} {21}},\ \bibinfo {pages} {661} (\bibinfo {year} {1967})}\BibitemShut {NoStop}%
\bibitem [{\citenamefont {Rogers}\ \emph {et~al.}(1969)\citenamefont {Rogers}, \citenamefont {Shannon}, \citenamefont {Sleight},\ and\ \citenamefont {Gillson}}]{rogers1969crystal}%
  \BibitemOpen
  \bibfield  {author} {\bibinfo {author} {\bibfnamefont {D.~B.}\ \bibnamefont {Rogers}}, \bibinfo {author} {\bibfnamefont {R.~D.}\ \bibnamefont {Shannon}}, \bibinfo {author} {\bibfnamefont {A.~W.}\ \bibnamefont {Sleight}},\ and\ \bibinfo {author} {\bibfnamefont {J.~L.}\ \bibnamefont {Gillson}},\ }\bibfield  {title} {\bibinfo {title} {Crystal chemistry of metal dioxides with rutile-related structures},\ }\href@noop {} {\bibfield  {journal} {\bibinfo  {journal} {Inorg. Chem.}\ }\textbf {\bibinfo {volume} {8}},\ \bibinfo {pages} {841} (\bibinfo {year} {1969})}\BibitemShut {NoStop}%
\bibitem [{\citenamefont {Cox}\ \emph {et~al.}(1982)\citenamefont {Cox}, \citenamefont {Cava}, \citenamefont {McWhan},\ and\ \citenamefont {Murphy}}]{cox1982neutron}%
  \BibitemOpen
  \bibfield  {author} {\bibinfo {author} {\bibfnamefont {D.~E.}\ \bibnamefont {Cox}}, \bibinfo {author} {\bibfnamefont {R.~J.}\ \bibnamefont {Cava}}, \bibinfo {author} {\bibfnamefont {D.}~\bibnamefont {McWhan}},\ and\ \bibinfo {author} {\bibfnamefont {D.}~\bibnamefont {Murphy}},\ }\bibfield  {title} {\bibinfo {title} {{A neutron powder diffraction study of the lithium insertion compound LiMoO$_2$ from 4--440K}},\ }\href@noop {} {\bibfield  {journal} {\bibinfo  {journal} {J. Phys. Chem. Solids}\ }\textbf {\bibinfo {volume} {43}},\ \bibinfo {pages} {657} (\bibinfo {year} {1982})}\BibitemShut {NoStop}%
\bibitem [{\citenamefont {Bolzan}\ \emph {et~al.}(1995)\citenamefont {Bolzan}, \citenamefont {Kennedy},\ and\ \citenamefont {Howard}}]{bolzan1995neutron}%
  \BibitemOpen
  \bibfield  {author} {\bibinfo {author} {\bibfnamefont {A.~A.}\ \bibnamefont {Bolzan}}, \bibinfo {author} {\bibfnamefont {B.~J.}\ \bibnamefont {Kennedy}},\ and\ \bibinfo {author} {\bibfnamefont {C.~J.}\ \bibnamefont {Howard}},\ }\bibfield  {title} {\bibinfo {title} {{Neutron powder diffraction study of molybdenum and tungsten dioxides}},\ }\href@noop {} {\bibfield  {journal} {\bibinfo  {journal} {Aust. J. Chem.}\ }\textbf {\bibinfo {volume} {48}},\ \bibinfo {pages} {1473} (\bibinfo {year} {1995})}\BibitemShut {NoStop}%
\bibitem [{\citenamefont {Leisegang}\ \emph {et~al.}(2005)\citenamefont {Leisegang}, \citenamefont {Levin}, \citenamefont {Walter},\ and\ \citenamefont {Meyer}}]{leisegang2005situ}%
  \BibitemOpen
  \bibfield  {author} {\bibinfo {author} {\bibfnamefont {T.}~\bibnamefont {Leisegang}}, \bibinfo {author} {\bibfnamefont {A.}~\bibnamefont {Levin}}, \bibinfo {author} {\bibfnamefont {J.}~\bibnamefont {Walter}},\ and\ \bibinfo {author} {\bibfnamefont {D.}~\bibnamefont {Meyer}},\ }\bibfield  {title} {\bibinfo {title} {{In situ X-ray analysis of MoO$_3$ reduction}},\ }\href@noop {} {\bibfield  {journal} {\bibinfo  {journal} {Cryst. Res. Technol.}\ }\textbf {\bibinfo {volume} {40}},\ \bibinfo {pages} {95} (\bibinfo {year} {2005})}\BibitemShut {NoStop}%
\end{thebibliography}%

\end{document}